\documentclass[
 reprint,
 amsmath,amssymb,
 aps,
]{revtex4-1}

\usepackage{graphicx}% Include figure files
\usepackage{dcolumn}% Align table columns on decimal point
\usepackage{bm}% bold math
\usepackage{hyperref}% add hypertext capabilities
\usepackage[dvipsnames]{xcolor}

\usepackage{tabularx}
\begin{document}

\preprint{APS/123-QED}

\title{Particle Production in AgAg Collisions at $E_{\rm Kin}=1.58A$ GeV within a Hadronic Transport Approach}% Force line breaks with \\

\author{Jan Staudenmaier$^{2,3}$, Natey K\"ubler$^{2,3}$, Hannah Elfner$^{1,2,3}$}

\address{$^1$ GSI Helmholtzzentrum f\"ur Schwerionenforschung, Planckstr. 1, 64291 Darmstadt, Germany}
\address{$^2$ Institut f\"ur Theoretische Physik, Johann Wolfgang Goethe-Universit\"at, Max-von-Laue-Str. 1, 60438 Frankfurt am Main, Germany}
\address{$^3$ Frankfurt Institute for Advanced Studies (FIAS), Ruth-Moufang-Stra{\ss}e 1, 60438 Frankfurt am Main, Germany}

\date{\today}% It is always \today, today,

\begin{abstract}
Heavy-ion collisions at low beam energies explore the high density regime of strongly-interacting matter. The dynamical evolution of these collisions can be successfully described by hadronic transport approaches. In March 2019, the HADES collaboration has taken data for AgAg collisions at $E_{\rm Kin}=1.58A$ GeV and in this work, we provide predictions for particle production and spectra within the Simulating Many Accelerated Strongly-interacting Hadrons (SMASH) approach. The multiplicities and spectra of strange and non-strange particles follow the expected trends as a function of system size.

In particular, in ArKCl (and pNb) collisions, much higher yields of double-strange baryons were observed experimentally than expected from a thermal model. Therefore, we incorporate a previously suggested mechanism to produce $\Xi$ baryons via rare decays of high mass $N^*$ resonances and predict the multiplicities. In addition, we predict the invariant mass spectrum for dilepton emission and explore the most important sources of dileptons above 1 GeV, that are expected to indicate the temperature of the medium. Interestingly, the overall dilepton emission is very similar to the one in AuAu collisions at $1.23 A$ GeV, a hint that the smaller system at a higher energy behaves very similar to the larger system at lower beam energy.
\end{abstract}

\pacs{Valid PACS appear here}% PACS, the Physics and Astronomy
\maketitle

%###############################################################################
\section{Introduction}
%###############################################################################

Studying the phase diagram of QCD, the fundamental field theory of the strong interaction, is one of the major goals of heavy-ion research. Heavy-ion collisions at low beam energies as explored by the HADES collaboration at GSI \cite{Agakishiev:2009am} allow to access the high density regime at low temperatures \cite{Cleymans:2005xv}. Since it is not obvious that the system reaches a state of (local) equilibrium, microscopic hadronic transport approaches have been successfully employed to describe the dynamics of such collisions \cite{Hartnack:1989sd,Bass:1998ca,Bratkovskaya:2007jk,Buss:2011mx,Weil:2016zrk}. Over the last 15 years, HADES has measured the hadron and dilepton production over a large variety of different collision systems (pp, pA and AA) at different collision energies.

The yields and spectra of different hadron species are the basic observables to study the
properties of hot and dense matter and the particle production mechanisms. Strange hadrons are of special interest, since their strangeness has
to be newly produced during the nuclear collision. The kinematic regime reached by the SIS-18 accelerator is close to the production threshold \cite{Adamczewski-Musch:2018xwg}, since strange particles are in general heavier than light hadrons. Therefore, secondary interactions are required and strangeness production is enhanced in nucleus-nucleus
reactions compared to elementary reactions \cite{Agakishiev:2009rr,Agakishiev:2010rs}. Complementary to the hadronic
observables is the measurement of dileptons. As an electromagnetic probe
dileptons escape the strongly-interacting medium unperturbed and allow the study of the matter over the whole lifetime of the reaction. The sensitivity
to the vector meson spectral function is expected to reveal the onset of chiral
symmetry in a hot and dense medium. In addition, the extraction of the
time-integrated temperature is in principle possible from the dilepton spectrum \cite{Adamczewski-Musch:2019byl}.

Transport models are successfully employed to reproduce hadron and dilepton production \cite{Bass:1998ca,Bratkovskaya:2007jk,Buss:2011mx,Weil:2016zrk,Hartnack:2011cn}, especially to
establish a baseline calculation based on vacuum resonance properties. Another
approach is the so-called thermal model, which is based on a (grand)-canonical
fit to experimental measurements of particle yields as for example realised in \cite{Wheaton:2004qb}. The agreement with data for
light and single strange hadrons to experimental data is surprisingly good for different
energies \cite{Andronic:2005yp}. However, for the few GeV energy regime discussed here discrepancies
appeared in the comparison to data for the $\Xi$ baryon and the $\phi$ meson, where much higher yields were observed than expected \cite{Agakishiev:2009rr}. There are several attempts to explain these high production yields \cite{Li:2012bga,Zetenyi:2018oyd,Steinheimer:2015sha}. 

In this work, the particle production is predicted for silver-silver (AgAg) collisions at $E_{\rm Kin}=1.58A$ GeV, which have been recently measured by the HADES collaboration. This is a new collision system with intermediate size at a slightly higher energy than the previously taken AuAu data at $E_{\rm Kin}= 1.23A$ GeV. The hadronic transport approach SMASH \cite{Weil:2016zrk} is employed to assess the expectations within a dense hadronic system, where the resonances follow vacuum Breit-Wigner spectral functions. In Section \ref{sec:model} the approach is described in more detail. Especially the production of $\Xi$ baryons from high mass resonances, similar in spirit to the work in \cite{Steinheimer:2015sha}, is explained. The predictions are based on a constraint of the decay probabilities from experimental data from elementary reactions, which is verified by comparisons to existing experimental data for ArKCl collisions. Comparison of those predictions with the upcoming data will allow to further constrain the viability of the $\Xi$ production from high mass resonances. A similar mechanism has already been successfully employed for the $\phi$ production in SMASH \cite{Steinberg:2018jvv}. Section \ref{sec:results} contains the predictions for particle multiplicities, rapidity and transverse momentum spectra of strange and non-strange particles. The system size dependence for strange particle production is explored. Estimates of the kinetic freeze-out temperature from the slopes of the transverse mass distributions are extracted. Additionally, the high quality dilepton measurement might allow to probe the temperature of the thermal medium by accessing the spectrum at invariant masses beyond the  $\phi$ peak. Therefore, the different lepton pair contributions for this region are investigated.

%###############################################################################
\section{Model Description} \label{sec:model}
%###############################################################################

The approach employed for the following results is a hadronic transport approach,
SMASH \cite{Weil:2016zrk} in the version 1.6 \cite{dmytro_oliinychenko_2019_3485108}. It is based on the relativistic Boltzmann equation. The
collision term for the few GeV energy regime is modeled by binary hadron
scatterings and excitation and decay of resonances. Those scatterings are governed by a
geometric collision criterion. The restriction to binary scatterings guarantees
detailed balance. Resonance properties are chosen according to their vacuum
properties and adjusted to fit elementary cross-sections for several reactions. The  partial width is treated as suggested by Manley and Saleski in~\cite{Manley:1992yb} (with  different parameters). No explicit in-medium modifications are incorporated
besides the  dynamically generated collisional broadening. The elementary cross
sections serve as the  main input for the approach and are constrained with
experimental data, where  possible. In general, isospin symmetry for particle properties and cross sections is assumed. The included
degrees of freedom are hadrons  and well-established hadronic resonances with a
mass up to $2.3$ GeV, which are  mostly based on the particle listing provided
by the PDG~\cite{Tanabashi:2018oca}. For an updated  list of the degrees of freedom see \cite{Steinberg:2018jvv} and for a more  comprehensive description of the
approach see \cite{Weil:2016zrk}.Note that this work employs SMASH in cascade mode (= no mean-field potentials) except for the results in Fig.~\ref{fig:pot_ratio}.

In addition to the hadronic degrees of freedom, SMASH includes the emission of
photons \cite{Schafer:2019edr} and dileptons \cite{Staudenmaier:2017vtq} perturbatively. 
Dileptons are produced either by direct or Dalitz decay of resonances ($\rho, \omega, \phi, \pi, \eta, \eta', \Delta$) as also explored with other established transport approaches (GiBUU~\cite{Weil:2012ji}, UrQMD~\cite{Schmidt:2008hm}, IQMD~\cite{Thomere:2007cj} and HSD~\cite{Bratkovskaya:2013vx}). Within SMASH, dilepton as well as hadron production has been extensively 
studied in the SIS energy regime ($E_{\rm Kin}=1-3A$ GeV) to study elementary, nucleon-nucleus and nucleus-nucleus systems with various sizes with good agreements with
experimental  data \cite{Weil:2016zrk}. In particular
relevant for the  following are the studies of strangeness and dilepton
production \cite{Steinberg:2018jvv,Staudenmaier:2017vtq}.

Also employed in this work is a coarse-graining approach for thermal dilepton
emission \cite{Endres:2015fna}, where macroscopic quantities are extracted locally from the
microscopic transport approach. This is achieved by splitting the microscopic
evolution of the system in space-time cells and averaging over many events. The
extracted quantities from those cells are baryon ($\rho_B$) and energy
($\epsilon$) densities. Employing an appropriate equation of state,
the temperature $T$ and the baryon chemical potential $\mu_B$ are extracted. Note that, while the equation of state assumes local thermal and chemical equilibration in the cells, non-equilibrium corrections are taken into account if the cells deviates from equilibrium \cite{Endres:2014zua}.Dileptons are emitted thermally from the cells utilizing rates based on medium modified spectral
functions \cite{Rapp:1999us, Rapp:2000pe}. The final result for the  dilepton production is a combination from the emission from the microscopic  transport and the thermal emission from the coarse-graining approach that is only applied for ("hot") cells where medium modifications are expected to play a role. For a general introduction into the employed coarse-graining approach the reader is referred to \cite{Endres:2015fna} and for a more detailed introduction of its application in the context of the SMASH transport approach to \cite{Staudenmaier:2017vtq}.

%-------------------------------------------------------------------------------
\subsection{$\Xi$ production}
%-------------------------------------------------------------------------------

Production of $\Xi$ baryons is famously underpredicted by theoretical approaches
\cite{Agakishiev:2010rs}. SMASH without an extension is no exception as can be seen in Table~\ref{tab:XiModelMulti} and \ref{tab:XiModelRatios} (results for SMASH-1.6). In SMASH, the $\Xi$ is produced by the decay of heavy hyperon resonances like
the $\Sigma(2030)$, $\Lambda(2100)$ or resonances of the $\Xi$ itself. They are
formed either by nucleon-kaon or hyperon-meson scatterings making the production
of $\Xi$ rare, since these are
 secondary scatterings with partners that are
often not abundant during a collision. In addition, the decaying resonances are
heavy and the branching ratios for decays involving the $\Xi$ are small, which
explains the underprediction for the $\Xi$ multiplicity.

Therefore, we follow the idea from \cite{Steinheimer:2015sha} and extend the approach by adding new decay channels for heavy $N^*$ resonances, namely $N^*\rightarrow\Xi KK$, in this
work. Comparing the $\Xi$ multiplicity obtained with this mechanism with the upcoming experimental data will show if these decays are a potential source of the seen $\Xi$ excess. This idea was already applied successfully for the $\phi$ meson
($N^*\rightarrow N \phi$) in \cite{Staudenmaier:2017vtq, Steinberg:2018jvv}.

However, the resonance treatment employed (\cite{Weil:2016zrk}, sec. II.C.3)
prevents adding any decays for which the combined pole mass of the final state
particles is larger than the pole mass from the decaying resonance, although
such a decay would be strictly physical speaking possible (see \cite{Steinberg:2018jvv}, sec.
II.A for a more detailed explanation). This restricts the addition of the new
decay channel to the heaviest $N^*$ resonances. The decay channel is added to
the two heaviest resonances ($N(2220)$, $N(2250)$) and their pole mass is
shifted slightly upwards ($N(2220)\rightarrow N(2350)$, $N(2250)\rightarrow
N(2400)$), which is possible due their relative large width ($ \Gamma \geq 400\,
\rm{MeV}$) and experimental uncertainties for the pole masses (on the order of
$100\,\rm{MeV}$).

\begingroup
\setlength{\tabcolsep}{5pt}
\renewcommand{\arraystretch}{1.1}
\begin{table}
	\begin{tabular}{ccc}
		\hline
		\hline \\[-1em]
		$\rm N(\Xi^-)$  & pNb & ArKCl \\
		\hline \\[-1em]
		SMASH-1.6& $\approx 0.0$ &$ 6.25\times10^{-7}$ \\
		$\rm \Xi^-$ from $\rm N^*$& $2.04\times 10^{-4}$ & $1.95\times 10^{-4}$ \\
		HADES & $(2.0 \pm 0.4 \pm 0.3) \times 10^{-4}$ &  $(2.3 \pm 0.9) \times 10^{-4}$ \\
		\hline
		\hline
	\end{tabular}
	\caption{\label{tab:XiModelMulti}$\rm \Xi^-$  yields in pNb ($E_{\rm Kin}=3.5$ GeV) and ArKCl ($E_{\rm Kin}=1.76A$ GeV) collisions compared to HADES results from \cite{Agakishiev:2015xsa, Agakishiev:2010rs}.}
\end{table}
\endgroup

\begingroup
\setlength{\tabcolsep}{5pt}
\renewcommand{\arraystretch}{1.1}
\begin{table}
	\begin{tabular}{ccc}
		\hline
		\hline \\[-1em]
		$\rm \Xi^-/(\Lambda+\Sigma^0)$ & pNb & ArKCl \\
		\hline \\[-1em]
		SMASH-1.6& $\approx 0.0$ & $1.1\times 10^{-5}$\\
		$\rm \Xi^-$ from $\rm N^*$ & $1.5\times 10^{-2}$ & $4.5\times10^{-3}$\\
		HADES &  $(1.2\pm 0.3\pm 0.4)\times 10^{-2}$ & $(5.6\pm1.2\pm ^{1.8}_{1.7})\times 10^{-3}$   \\
		\hline
		\hline
	\end{tabular}
	\caption{\label{tab:XiModelRatios}$\rm \Xi^-/(\Lambda+\Sigma^0)$ ratios in pNb ($E_{\rm Kin}=3.5$ GeV) and ArKCl ($E_{\rm Kin}=1.76A$ GeV) collisions compared to HADES results from \cite{Agakishiev:2015xsa, Agakishiev:2010rs}.}
\end{table}
\endgroup

The branching ratio for new $N^*\rightarrow\Xi KK$ is now constrained with the
experimental data from pNb reactions, which is the most elementary data
available. We find $BR(N^*\rightarrow\Xi KK)=0.5$. The results for pNb before and after addition of the $\Xi$ production from heavy $N^*$ decays is shown in Table~\ref{tab:XiModelMulti}. Before, although theoretical possible, no $\Xi$ production (for the given number of calculated reactions) is observed due to the lack of many secondary reactions in the small pNb system. After the addition, with the tuned $BR(N^*\rightarrow\Xi KK)$ of $0.5$ the experimental data from HADES \cite{Agakishiev:2015xsa} can be matched. The same is true for the $\rm \Xi^-/(\Lambda+\Sigma^0)$ ratio as seen in Table~\ref{tab:XiModelRatios}.

Compared to \cite{Steinheimer:2015sha} a larger branching ratio is reported here, since the decay is included for less and heavier resonances. A similar observation was made for the branchings of $N^*$ decays into $\phi$ in \cite{Steinberg:2018jvv, Staudenmaier:2017vtq}. Interestingly, if one calculates the ratio of the two branching ratios of $\phi$ and $\Xi$, $BR(N^*\rightarrow\phi N)/(BR(N^*\rightarrow\Xi KK)=0.02$, the relation exactly matches for the two branching ratios reported in \cite{Steinheimer:2015sha}. (Note that $BR(N^*\rightarrow\phi N)$ is updated to 0.01 in comparison to \cite{Steinberg:2018jvv, Staudenmaier:2017vtq} to account for new experimental constraints e.g. by the updated PDG~\cite{Tanabashi:2018oca}.)

Next, the $\Xi$ production is compared to available
experimental data for a larger system to verify the newly introduced treatment. Tables~\ref{tab:XiModelMulti} and \ref{tab:XiModelRatios} show that the $\Xi$ production in ArKCl is also matched well within the errors. In the following, the calculations with the adjusted branching ratios are dubbed ''modified branching ratios', otherwise all calculations are performed within the default SMASH settings. On this basis, predictions for the $\Xi$ production for AgAg collisions are presented below (in section~\ref{sec:multi}).

%###############################################################################
\section{Results} \label{sec:results}
%###############################################################################

In the following, the predictions from SMASH-1.6 are shown for several hadronic and electromagnetic observables. For the relevant observables, the effect of the additional $N^*$ decay channels for the $\Xi$ production is discussed.

%-------------------------------------------------------------------------------
\subsection{Multiplicities} \label{sec:multi}
%-------------------------------------------------------------------------------

\begin{figure}
\includegraphics[width=0.95\columnwidth]{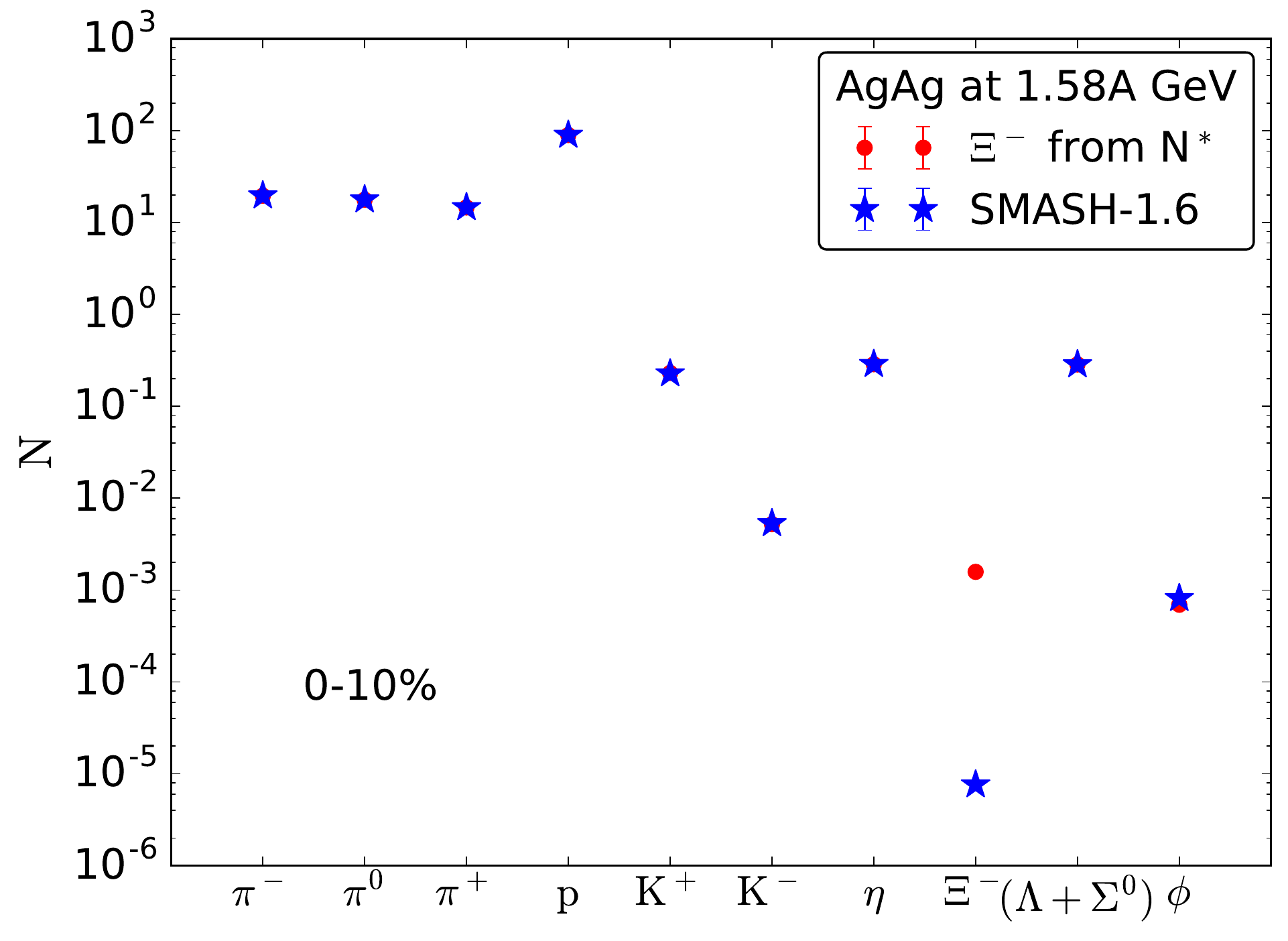}\caption{\label{fig:multies}Results for the particle production yields for AgAg collisions at $E_{\rm Kin}=1.58A$ GeV at 0-10\% centrality from SMASH-1.6 (stars) compared to the modified branching ratios for high mass resonances (circles).}
\end{figure}

Let us start with the overall total multiplicities. Predictions for the particle yields in AgAg collisions at $E_{\rm Kin}=1.58A$ GeV in the $0-10$\% centrality class are depicted in Fig. \ref{fig:multies}. Here, the results from SMASH-1.6 (stars) are confronted with the yields resulting from the version including additional $N^*\to\Xi KK$ resonance decays (circles).

Clearly, mostly protons and pions are produced while there is a hierarchy in the production of the latter with a charged pion ratio $\pi^+/\pi^-<1$ according to the isospin imbalance in AgAg collisions. All strange particles show a significantly lower production rate due to their higher masses and the fact that strange quarks need to be newly produced. 

\begingroup
\setlength{\tabcolsep}{10pt} 
\renewcommand{\arraystretch}{1.1} 
\begin{table}[h]
	\begin{tabular}{ccc}
		\hline
		\hline \\[-1em]
		AgAg & $\rm N(\Xi^-)$ & $\Xi^-/(\Lambda+\Sigma^0)$\\
		\hline \\[-1em]
		SMASH-1.6 & $8.50\times 10^{-6}$ & $2.71\times 10^{-5}$\\
		$\rm \Xi^-$ from $\rm N^*$ & $1.78\times 10^{-3}$ & $5.63\times10^{-3}$\\
		\hline
		\hline
	\end{tabular}
	\caption{\label{tab:XiAgAg}Predictions for $\rm \Xi^-$  yield and $\Xi^-/(\Lambda+\Sigma^0)$ ratio in AgAg collisions.}
\end{table}
\endgroup

The addition of the resonances decaying into $\Xi^-$-baryons only influences the $\Xi^-$ yield itself significantly. The predicted values for average multiplicity and $\rm \Xi^-/(\Lambda+\Sigma^0)$ ratios are also explicitly given for future reference in Table~\ref{tab:XiAgAg}. Since the $\Xi^-$ is produced in a multi-step process similar to $K^+$~\cite{Hartnack:2011cn}, the yield is sensitive to the employed treatment of the Fermi motion, the resonance lifetime and the underlying equation of state for the potentials, which influences the density near the collision center and in term also the resonance lifetimes. Here, the same treatment established in earlier studies with SMASH~\cite{Weil:2016zrk,Staudenmaier:2017vtq,Steinberg:2019wgm,Steinberg:2018jvv} without potentials is chosen to limit the uncertainties and make the calculation numerically less expensive. The $\phi$ meson yield is slightly lowered, since the addition of a new decay channel lowers the weight of the existing channels. The effect is fortunately small, since the $\Xi^-$ is merely produced once in 1000 collisions while the production rates of the other particles are significantly higher. Although the change in all other particle yields is small, we stick to SMASH-1.6 without additional decays of heavy nucleon resonances into $\Xi$ baryons in the following, unless explicitly stated otherwise.

At the collision energy of a few GeV per nucleon, it is important to think about the definition of participants and spectators since they mix in phase-space and no easy separation by kinematic cuts is possible. We present the influence of different selection cuts in Appendix \ref{appendix_spec}. Only protons that interact either elastically or inelastically are included in Fig. \ref{fig:multies} and with this all spectators are cut out. Furthermore, the collisions have been divided into centrality classes according to their impact parameter, following the results of Glauber calculations presented in \cite{Miskowiec}. The centrality classes, their respective impact parameters and numbers of participants for AuAu, AgAg, ArKCl and CC collisions are summarized in Appendix \ref{appendix_centrality}. For comparison, the experimental centrality determination of HADES is presented in detail in \cite{Adamczewski-Musch:2017sdk}.

When nucleons surpass a critical distance in phase space, they may form bound states and produce clusters. These clustering effects are not considered here. Approaches to include these effects are usually based on effectively fine tuning the critical phase space parameters to pre-existing experimental data which we refrain from doing here, since the experimental data is not available yet.

\begin{figure}
	\includegraphics[width=0.95\columnwidth]{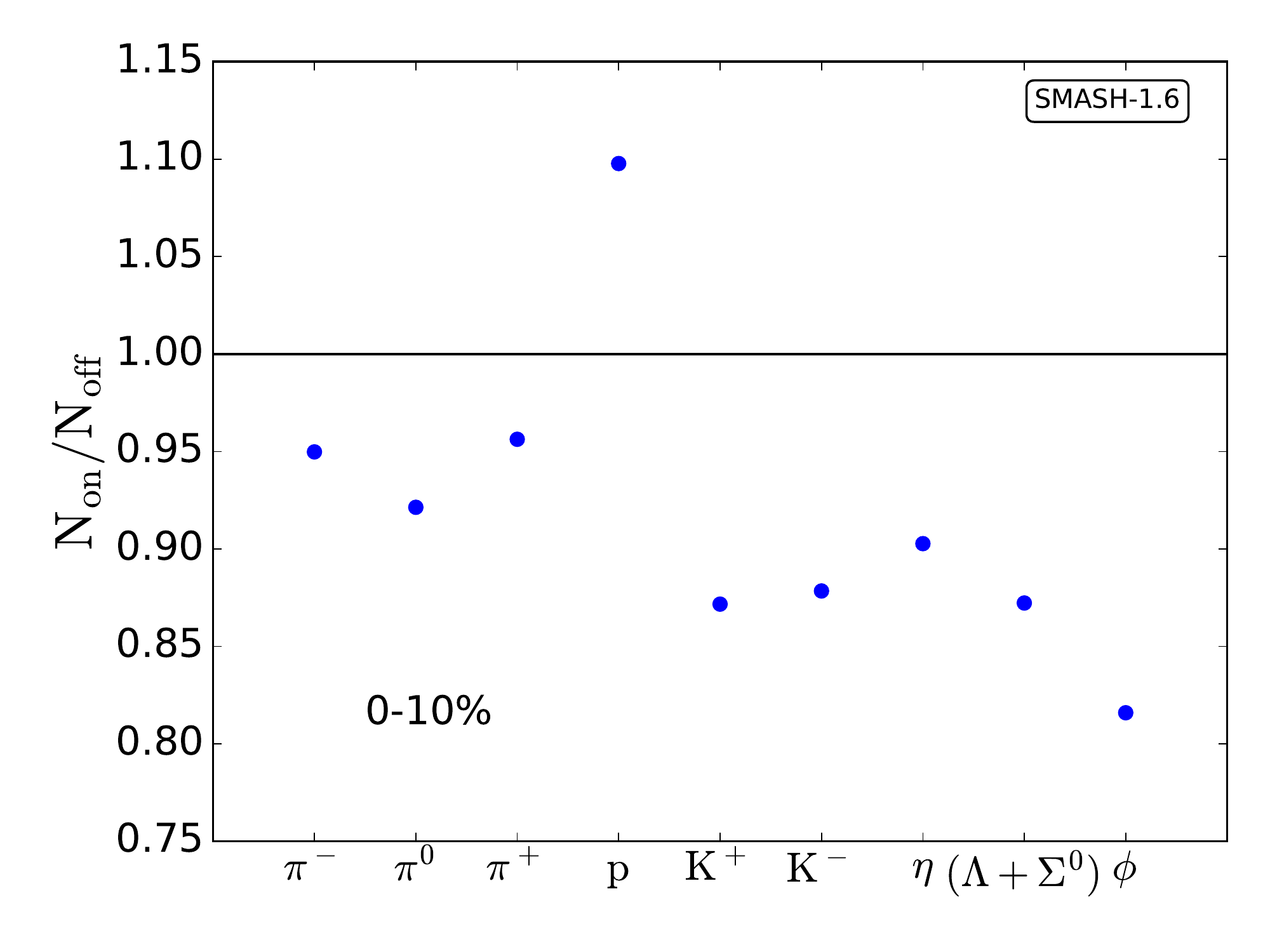}\caption{\label{fig:pot_ratio}Ratio of particle yields with (on) and without potentials (off) in AgAg collisions at $E_{\rm Kin}=1.58A$ GeV.}
\end{figure}

To understand the effects of nuclear mean fields the ratio of the SMASH results with and without potentials in central AgAg collisions at $E_{\rm Kin}=1.58A$ GeV has been calculated in Fig. \ref{fig:pot_ratio} for all particle species (except the $\Xi$ baryon). In the current calculation, a simple density dependent Skyrme potential has been included with parameters corresponding to a compressibility of $\kappa =240$ MeV \cite{Xu:2016lue}. While the proton yields are enhanced all other species are suppressed, when mean field interactions are taken into account. The effect on pions and $\eta$ mesons is smaller as expected.  Overall the multiplicities can change between 5-25 \%, when mean fields are taken into account.

%-------------------------------------------------------------------------------
\subsection{Rapidity Spectra}
%-------------------------------------------------------------------------------
\begin{figure}

	\includegraphics[width=0.95\columnwidth]{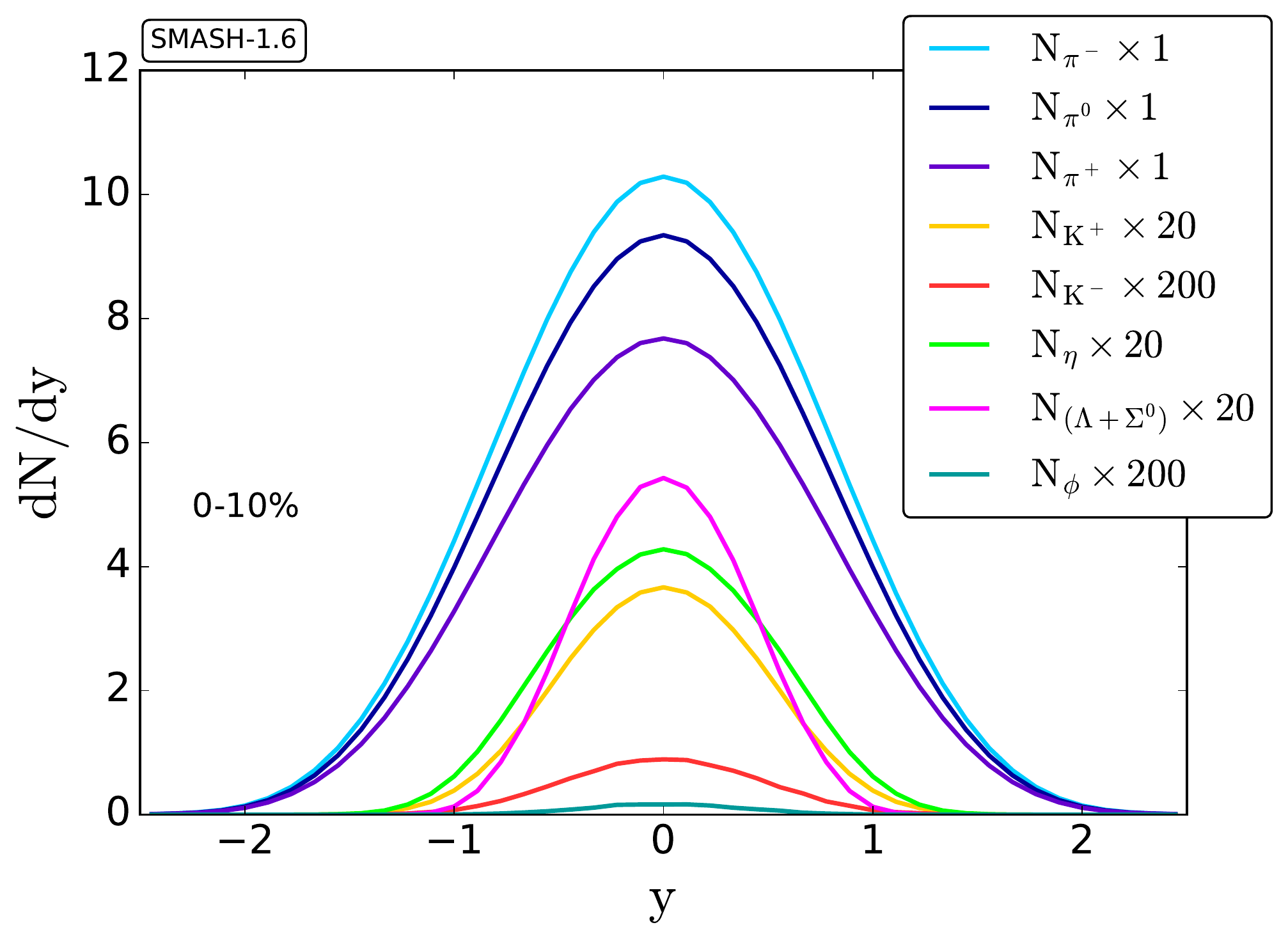}
	\includegraphics[width=0.95\columnwidth]{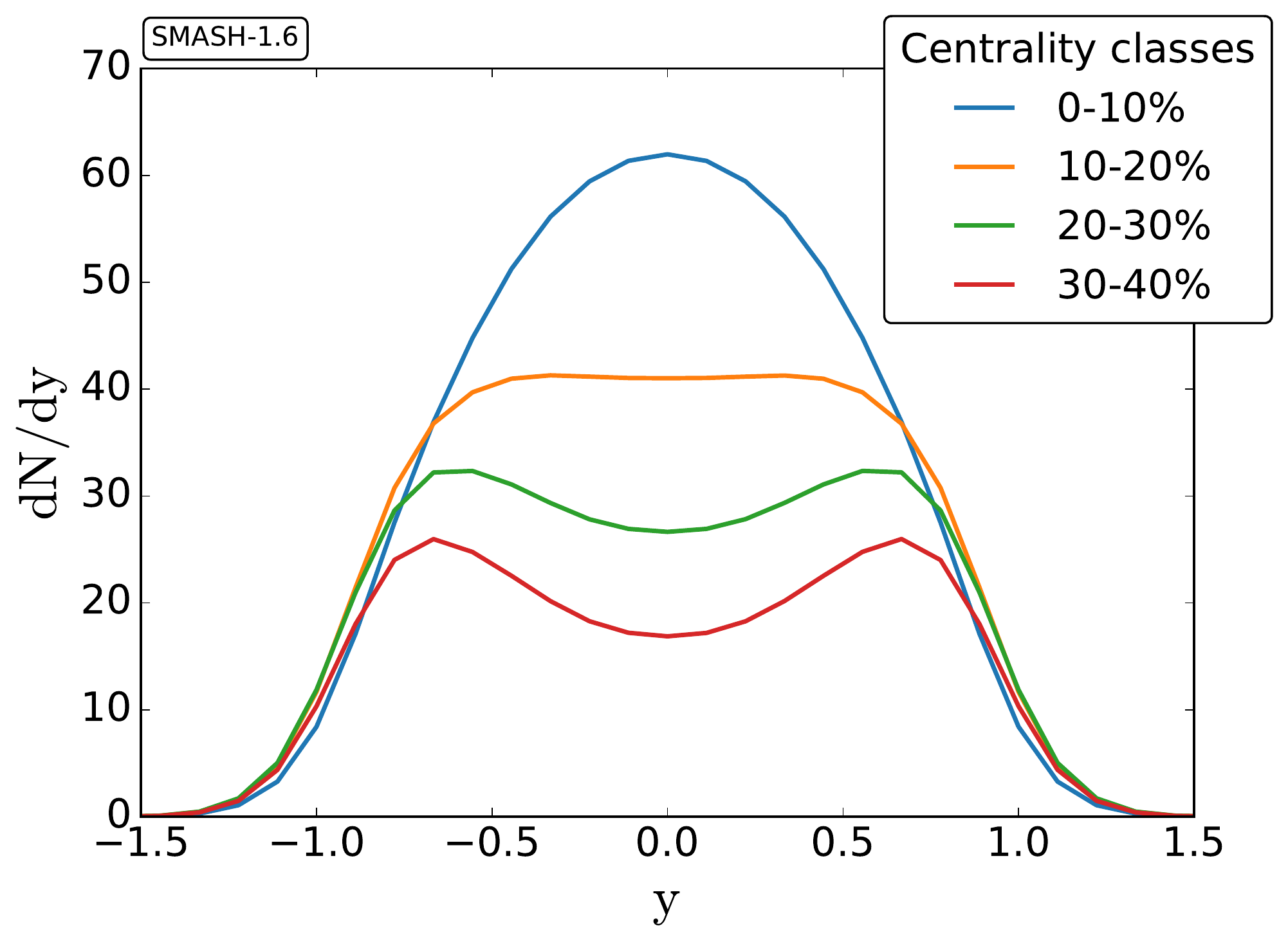}
	\caption{Rapidity distributions of $K^{\pm}$, $\Lambda$, $\phi$ and $\eta$ at 0-10\% centrality (top) and of protons in different centrality classes (bottom) in AgAg collisions at $E_{\rm Kin}=1.58A$ GeV.}
	\label{fig:y_all}
\end{figure}

Let us turn next to more differential distributions of particle production. The rapidity spectra contain information on the longitudinal dynamics of particle production and Fig. \ref{fig:y_all} (top) shows the ones for pions, kaons, $\eta$ and $\phi$ mesons as well as $\Lambda$ baryons  in central collisions. All strange particles have been scaled up by factors of 20 or 200 to be visible and distinguishable. The production hierarchy already observed in the multiplicities in Fig. \ref{fig:multies} is nicely reproduced. As expected, the pions have the largest yields, where again the isospin asymmetry matches the one of the collision system, the $\pi^-$ production visibly outweighs the $\pi^+$ yields. 

All newly produced particles follow a Gaussian shape as a function of rapidity as expected. The difference between the $K^+$ and $K^-$ yields comes from the fact that the due to the quark content of the $NN$ scatterings, the $K^-$-meson is always produced together with $K^+$ via $NN \rightarrow NNK^+ K^-$, whereas $K^+$ is also produced alone via $NN \rightarrow \Lambda NK^+$. The rapidity distributions for protons in four centrality classes between $0$\% and $40$\% in AgAg collisions at E$_{\rm Kin}=1.58A$ GeV are summarised in Fig. \ref{fig:y_all} (bottom). The yields decrease with increasing centrality classes and a drop around the mid-rapidity region becomes more pronounced. From these rapidity distributions, we conclude that our selection of participants and the general dynamics in central, mid-central and peripheral collisions works as expected within SMASH-1.6. 

%-------------------------------------------------------------------------------
\subsection{Transverse Mass Spectra}
%-------------------------------------------------------------------------------

\begin{figure}
	\includegraphics[width=0.95\columnwidth]{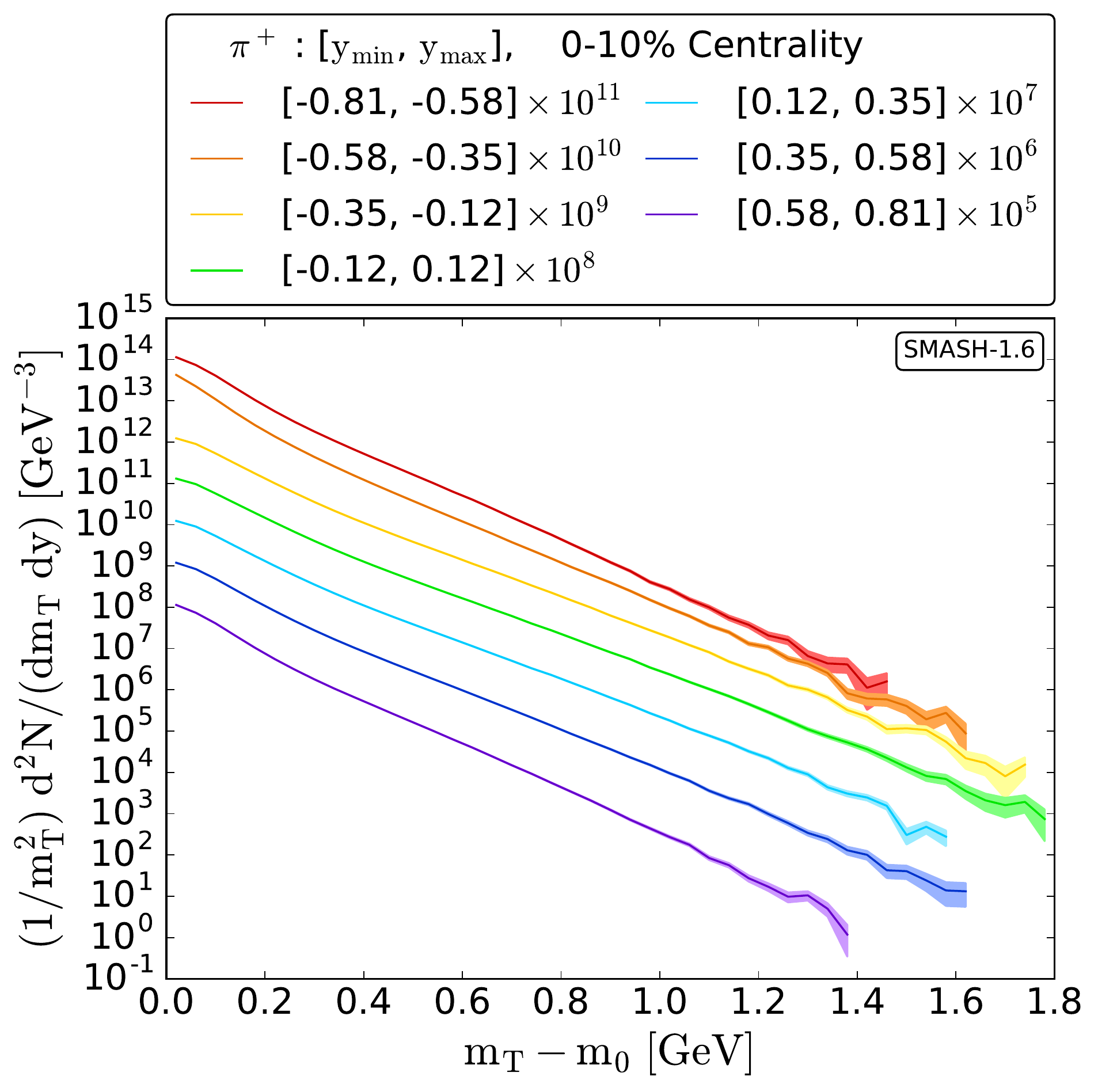}
	\caption{\label{fig:m_t_rare}Mean transverse mass distribution of the $\pi^+$ meson in AgAg collisions at $E_{\rm Kin}=1.58A$ GeV. }
\end{figure}

\begin{figure}
	\includegraphics[width=0.95\columnwidth]{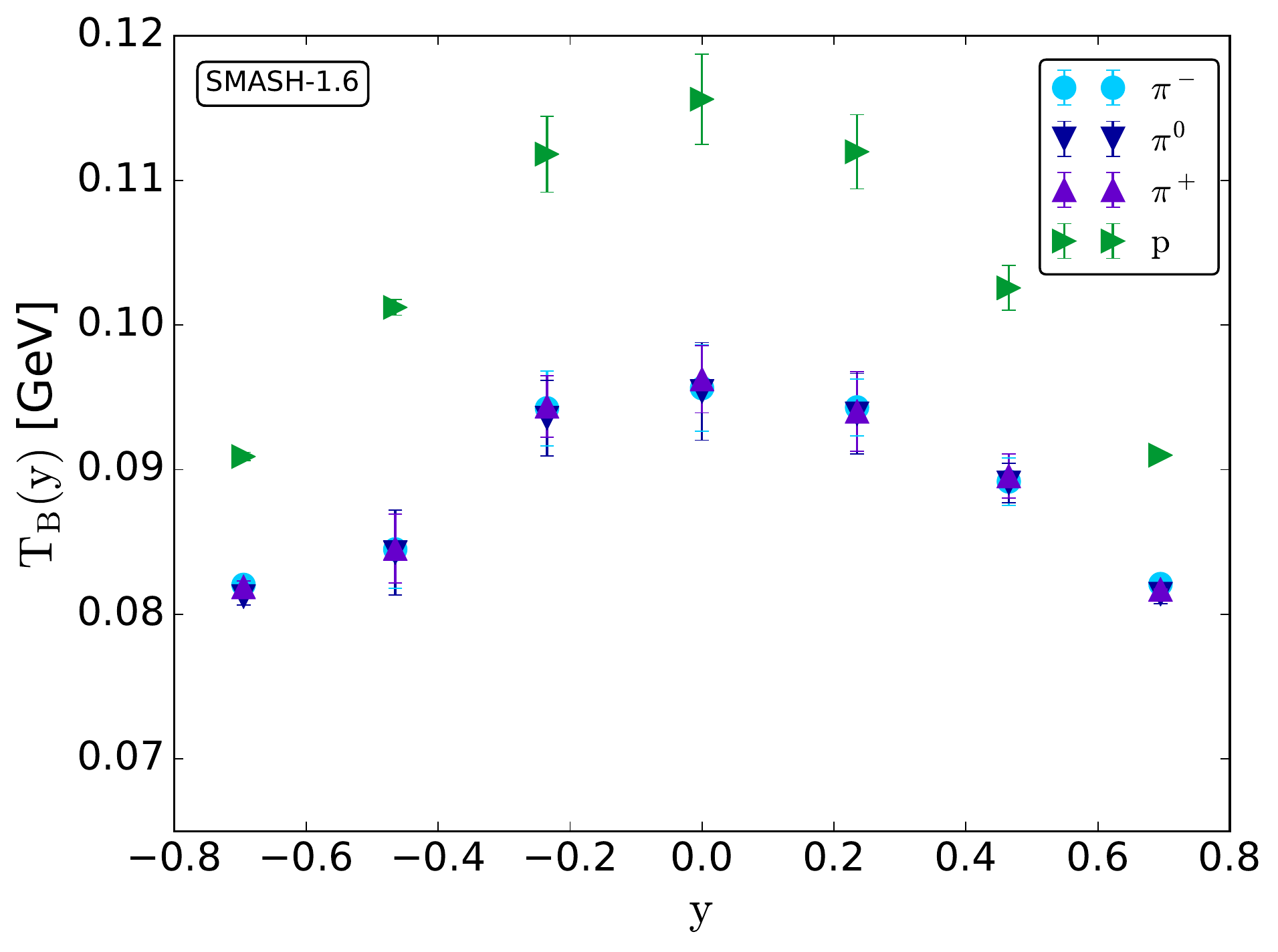}
	\includegraphics[width=0.95\columnwidth]{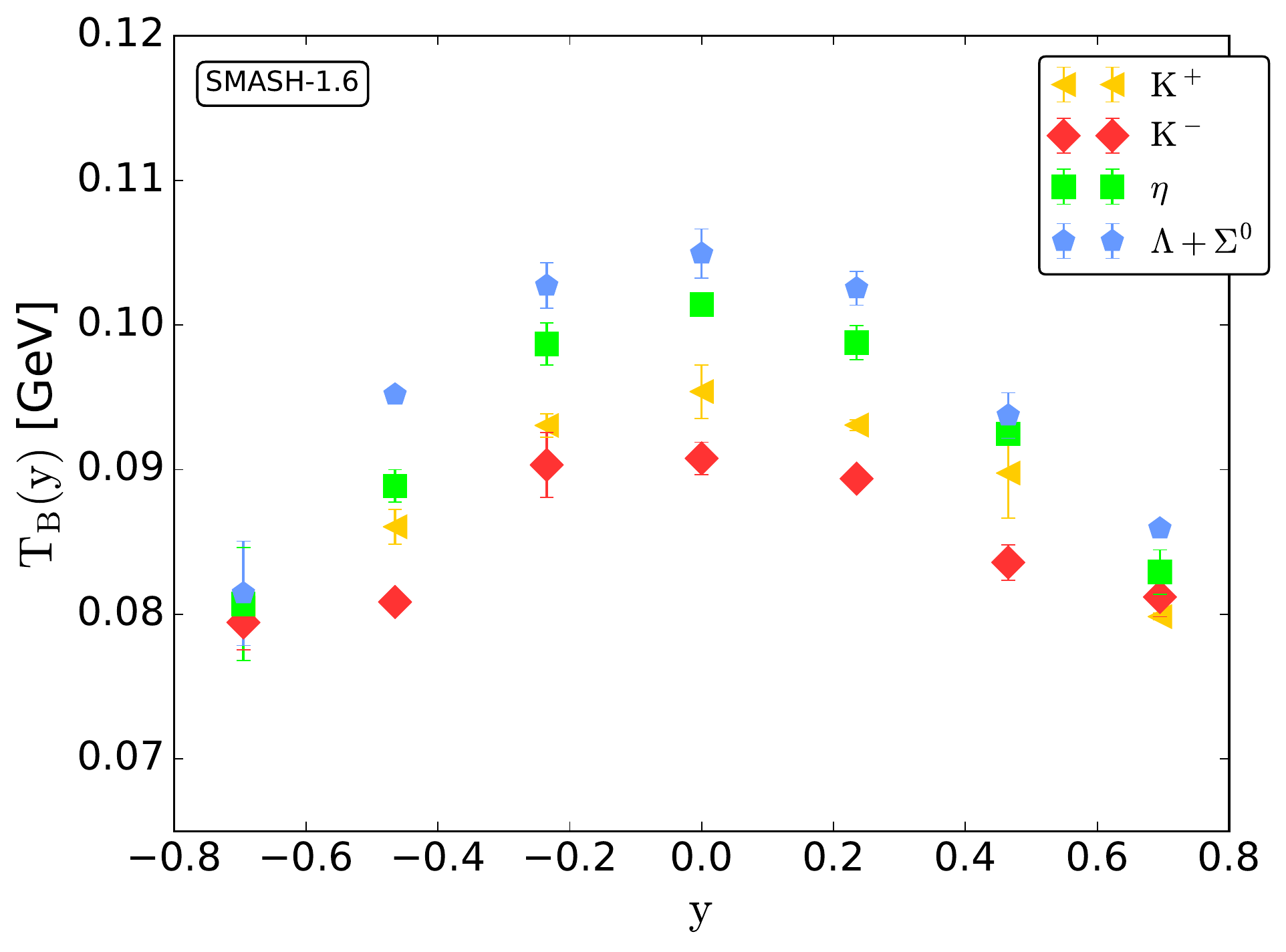}
	\caption{\label{im:Temperature}Inverse slope parameter $T(y)$ as a function of rapidity $y$ for protons and pions (top) and strange particles (bottom) AgAg collisions at E$_{\rm Kin}=1.58A$ GeV.}
\end{figure}

Fig. \ref{fig:m_t_rare} depicts the transverse mass spectra of positively charged pions for several rapidity bins as an example case. While we choose to discuss more inclusive information on effective temperatures extracted from spectra in the following, the detailed spectra for all other species are shown in Appendix \ref{appendix_transverse_mass}. Note that the modified $\Xi^-$ production channels do not interfere with the (mean) transverse masses for neither the $\Xi^-$ baryon itself nor for the other particles.  A Boltzmann fit of the form
\begin{equation}\label{eq:Boltzmann}
\frac{1}{m_T^2}\frac{d^2N}{dm_Tdy}=C(y)\exp\left(-\frac{m_T-m_0}{T_B(y)}\right)
\end{equation}
with the rapidity dependent inverse slope parameter $T_B(y)$ and a likewise rapidity dependent normalization constant $C(y)$ is applied for all particle species and the extracted slope parameters as a function of rapidity are shown in Fig. \ref{im:Temperature}. The error bands in Fig. \ref{fig:m_t_rare} include the statistical errors only. At high transverse mass the errors increase significantly and the spectra have been cut when the uncertainties become as large as the calculated value itself. Hence, the fitting region varies strongly depending on both the rapidity bin itself and also the particle species.

Under the assumption of a thermal source the inverse slope parameter $T_B$
is connected to the effective temperature $T_{\rm eff}$, defined as

\begin{equation}\label{eq:T_eff}
T_B=\frac{T_{\rm eff}}{\cosh(y)}.
\end{equation}
By fitting the inverse slope parameters $T_B$ in Fig. \ref{im:Temperature}
according to equation \ref{eq:T_eff} the effective temperatures $T_{\rm eff}$ in Table \ref{tab:T_eff} are obtained. In Fig. \ref{im:Temperature} the extracted inverse slope parameters $T_B$ as a function of rapidity $y$ for various particle species in central collisions are presented. 

The transverse mass distribution cannot be fitted with a single exponential. In order to obtain $T_B$, we define an upper and lower region of the distributions in which two independent exponential fits according to \eqref{eq:Boltzmann} are applied, and calculate the mean of these two values. Since the spectra are cut when the statistical uncertainties reach 100\%, we define the maximum value of each spectrum as the last bin with sufficient statistics. For all particles except the $K^-$ meson the upper region was chosen between 0.4 GeV and 60\% of the maximum value and the lower region between 0.2 GeV and 35\% of the maximum value for each rapidity bin. For the $\pi ^+$, for example, the fit in the rapidity bin [0.58,0.81] is performed in the ranges [0.2 GeV,0.641 GeV] and [0.4 GeV,1.036 GeV]. Due to the large statistical uncertainties the $K^-$ meson was fitted individually in the ranges [0.19 GeV,2.2 GeV] and [0.25 GeV,0.35 GeV].The error bars reflect the deviances of the two fits from the obtained mean.

\begin{table*}
	\centering
	\begin{tabular}{c|cccccccc}
		\hline
		\hline
		& $\pi^-$ & $\pi^0$ & $\pi^+$ & $p$ & $K^+$ & $K^-$ & $\eta$ & $\Lambda$ \\
		\hline
		$\rm T_{eff}(y)$ [MeV] & 98.0 & 97.5 & 98.1 & 114.2& 97.4 & 93.9&
       101.4& 105.0\\
		\hline
		\hline
	\end{tabular}
	\caption{\label{tab:T_eff}Effective temperatures $T_{\rm eff}$ for various
		particles in AgAg collisions at $1.58A$ GeV for 0-10\% centrality.}
\end{table*}

All results show the expected rapidity dependence with higher inverse slopes, harder spectra, at mid-rapidity and steeper spectra at forward and backward rapidities reflecting the kinematically available momenta. In addition, the values for the slopes for all particle species are very similar and in the expected range of 90-110 MeV. 

%-------------------------------------------------------------------------------
\subsection{System Size Dependence}
%-------------------------------------------------------------------------------

To study the effects of system size and put our results into context, the AgAg results are confronted with results from the smaller collisions systems CC and ArKCl, and the much larger system AuAu, all of which have previously been studied as part of the HADES experimental program at lower beam energies of $1.00$-$1.23A$ GeV. Since only the size of the colliding nuclei is of interest and any other energy related effects must be excluded, all collisions are calculated using the same energy of $E_{\rm Kin}=1.58A$ GeV in SMASH.

In \cite{Adamczewski-Musch:2018xwg}, Adamczewski-Musch \textit{et al.} show that in AuAu collisions at $E_{\rm Kin}=1.23A$ GeV the strange particles $\Lambda$, $K^{\pm, 0}$ and $\phi$ scale with the same constant $\alpha$ when the multiplicities $N$ are plotted as a function of the number of participants $N_{\rm part}$ and normalized by $N_{\rm part}$. If the only difference between the different collision systems would be their geometrical size, there would be a constant behaviour expected, since the system size scales with the number of participants. In \cite{Adamczewski-Musch:2018xwg} the HADES collaboration quotes that $\alpha$ describes the extra energy for particle production provided by the system.

\begin{figure}
	\includegraphics[width=0.95\columnwidth]{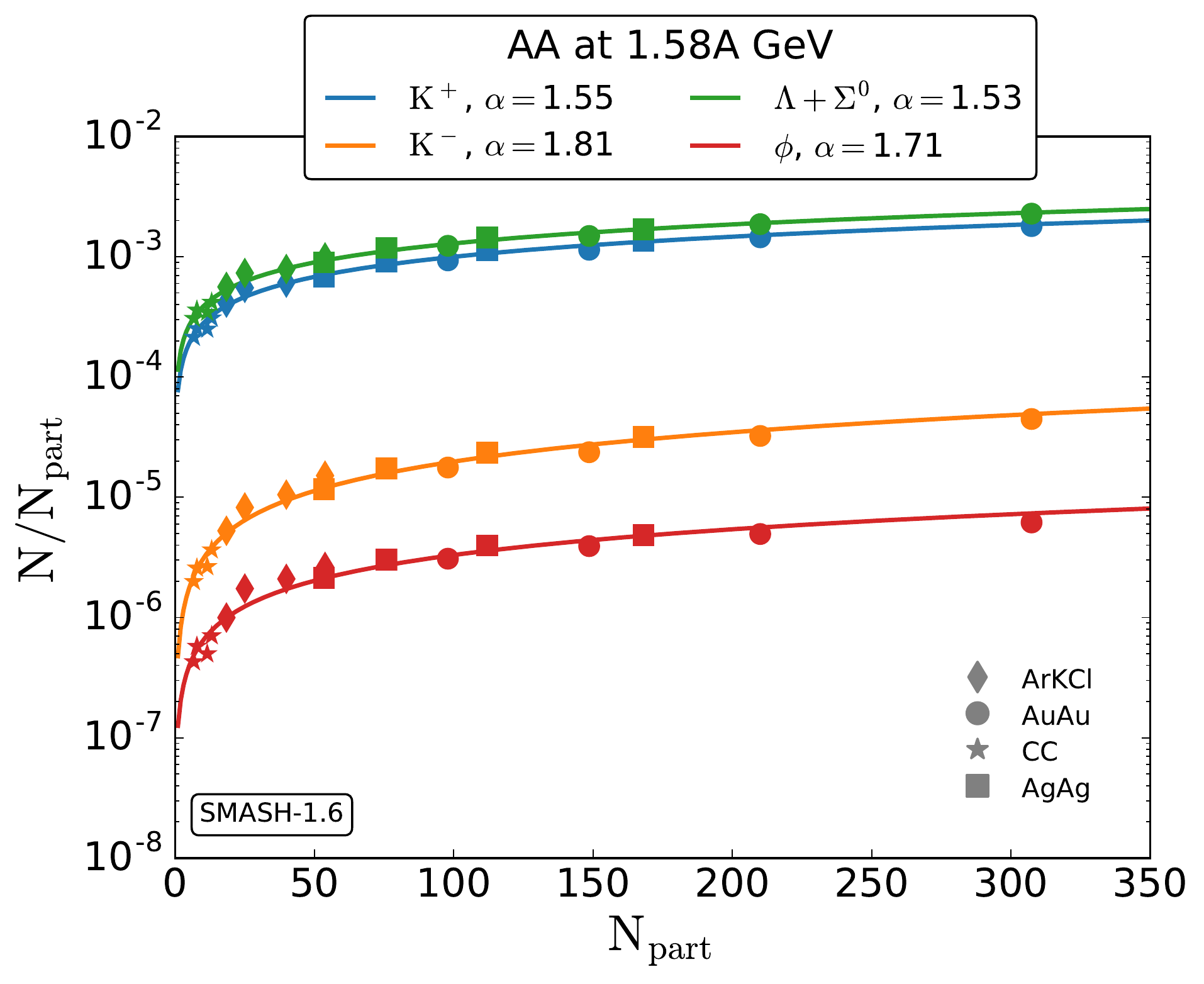}
	\caption{\label{fig:N_part}Multiplicities  per  number of participants $\rm N/N_{part}$ as  a  function  of $\rm N_{part}$.  Each hadron yield is fitted individually with a function $\rm N\propto (N_{part})^{\alpha}$.}
\end{figure}

We investigate here the behaviour of $\Lambda$, $K^{\pm}$ and $\phi$ in four different collision systems CC (stars), ArKCl (diamonds), AgAg (squares) and AuAu (circles) at $E_{\rm Kin}=1.58A$ GeV. The multiplicities from each corresponding centrality class are plotted against the number of participants and, using the method of least squares, the results are fitted according to

\begin{align}\label{eq:Npart_scaling}
\frac{N}{N_{\rm part}} &= C\cdot \left(N_{\rm part}\right)^{\alpha-1}=C\cdot e^{(\alpha -1) \cdot \ln (N_{\rm part})}\nonumber\\
\Leftrightarrow\quad \ln \left(\frac{N}{N_{\rm part}}\right) &= \ln (C) + (\alpha -1)\cdot \ln (N_{\rm part}) 
\end{align}

as shown by the lines in Fig. \ref{fig:N_part}. The impact parameters and numbers of participants for each centrality class are determined using Glauber calculations provided by D. Miskowiec in \cite{Miskowiec} and can be found in Table \ref{tab:c_cent} in the Appendix \ref{appendix_centrality}.
The $K^+$ meson and $(\Lambda +\Sigma^0)$ baryons scale, in fact, with similar values of $\alpha_{K^+}=1.55\pm 0.02$ and $\alpha_{\Lambda +\Sigma^0}=1.53\pm 0.02$. In particular, $\alpha_{\Lambda +\Sigma^0}$ is in agreement with the experimental result in \cite{Adamczewski-Musch:2018xwg} of $\alpha^{\rm exp}=1.45\pm0.06$ while the result for the $K^+$ mesons is very close to agreement.  The $\phi$ meson scales with $\alpha_{\phi}=1.71\pm 0.03$ and the $K^-$ with $\alpha_{K^-}=1.81\pm 0.03$, deviating by less than 20\% from the experimental value. 
A possible explanation for these deviations is that the collisions have been performed at a higher beam energy which leads to a higher disposition of energy within the systems. While using the centrality classes provided by \cite{Miskowiec} ensures compatibility with the upcoming experimental results for the centrality classes, the $N_{\rm part}$ value by the Glauber is model-dependent and therefore introduces an uncertainty of the $\alpha$ exponents that could also explain deviations. Even though, testing the $\alpha$ fit with $N_{\rm part}$ values from \cite{Adamczewski-Musch:2017sdk} for AuAu, no difference is found within errors.
Furthermore, the $\alpha $ parameter for $K$ mesons is sensitive to the stiffness of the EoS, when potentials are employed, as Hartnack \textit{et al.} found in \cite{Hartnack:2011cn}.
Lastly, we note that the chosen fitting procedure may lead to slight variations in the $\alpha$ constants.

%-------------------------------------------------------------------------------
\subsection{Dileptons}
%-------------------------------------------------------------------------------

\begin{figure} 
	\includegraphics[width=0.95\columnwidth]{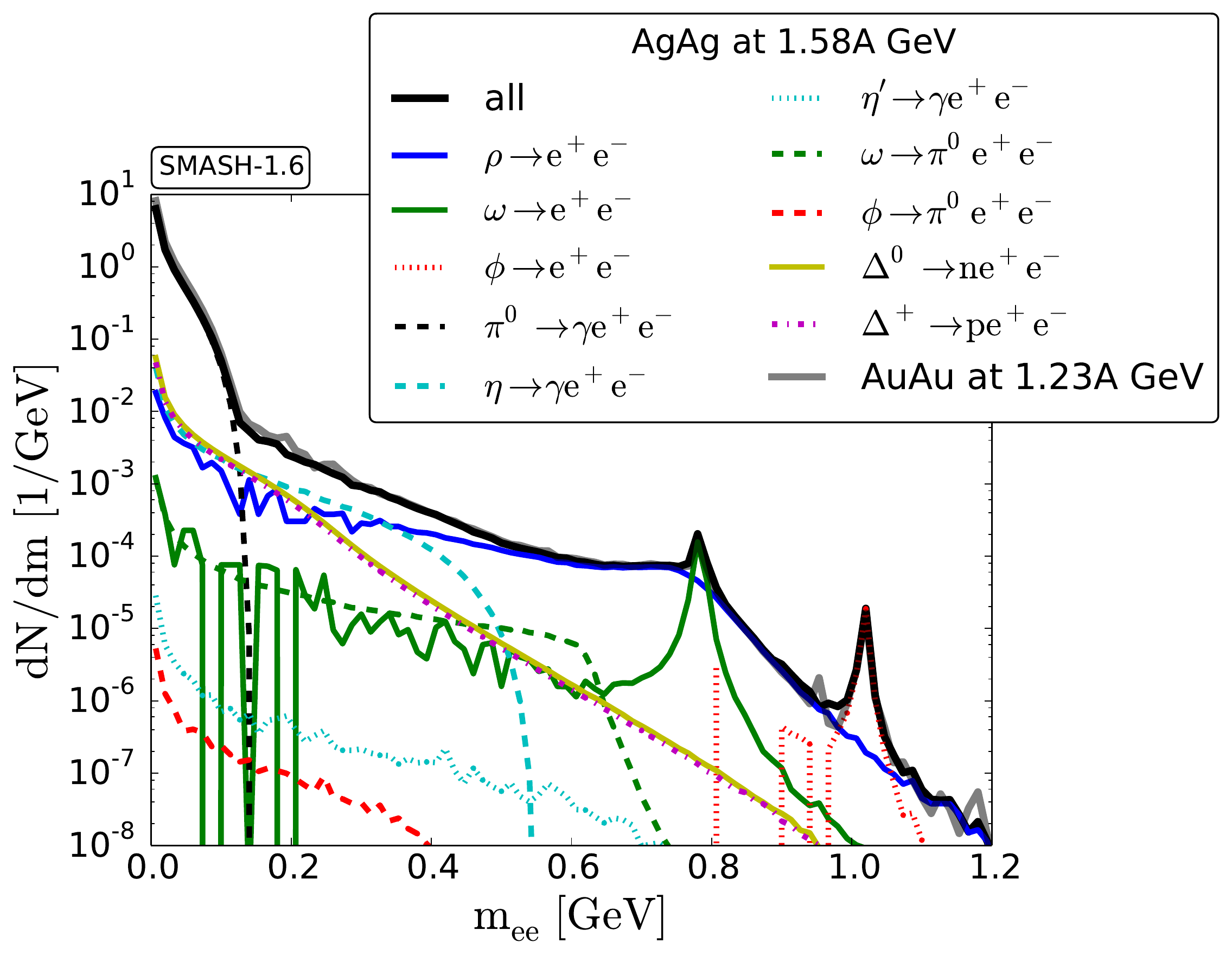}
	\caption{
		\label{fig:dil-AgAg}
		Invariant mass spectrum of dielectrons produced by AgAg collisions at $E_{\rm{Kin}}=1.58A\,\textrm{GeV}$.
	}
\end{figure}
The study of dilepton production complements the hadronic observables. The comparison between different collision systems allows to assess the magnitude of medium effects. Fig.~\ref{fig:dil-AgAg} shows the invariant mass spectrum for dielectrons produced in AgAg collisions for a kinetic energy of $1.58A$ GeV. The different channels contributing to the spectrum are displayed as well with dominant contributions from the $\pi^0$ decay for low masses and the vector meson decays ($\rho$, $\omega$ and $\phi$) around their respective pole masses. 

Also shown in Fig.~\ref{fig:dil-AgAg} is the yield of the other large collision system studied by HADES, which is AuAu at $E_{\rm{Kin}}=1.23A\,\textrm{GeV}$. The total yield of the larger system AuAu and lower energy is strikingly similar. Only a slightly higher yield is observed for smaller masses, the higher beam energy of AgAg seems to overall compensate for the smaller system concerning the dilepton production.

\begin{figure} 
	\includegraphics[width=0.95\columnwidth]{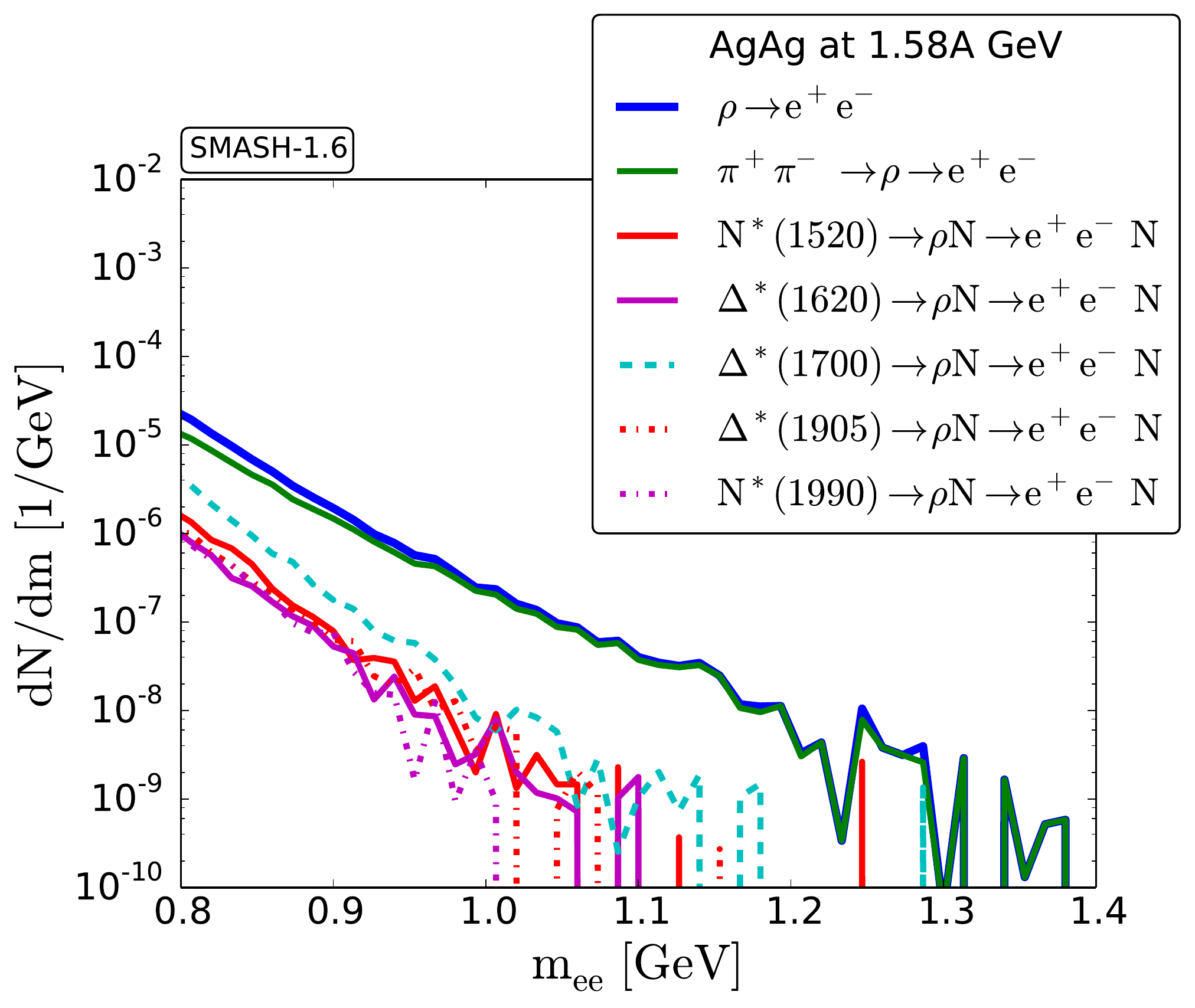}
	\caption{
		\label{fig:dil-AgAgrho}
		Contributions to the high mass tail of the $\rho\rightarrow e^+e^-$ decay in the invariant mass spectrum of dielectrons produced by AgAg collisions at $E_{\rm{Kin}}=1.58A\,\textrm{GeV}$.
	}
\end{figure}

Of special interest in the invariant mass spectrum is the yield above the $\phi$ peak, which  offers insights into the temperature of the medium and might become experimentally accessible for the first time with the upcoming high-statistics data for AgAg at these low energies. Therefore, it is important to understand all dilepton emitting sources in this region of the spectrum.

As seen in Fig.~\ref{fig:dil-AgAg}, the dilepton production observed with SMASH above the $\phi$ peak is dominated by the $\rho$ contribution. Fig.~\ref{fig:dil-AgAgrho} therefore shows the different processes from which the $\rho$ yield originates in the mass region above and around the $\phi$ pole mass. The $\rho$ that decays into the electron pair is either produced by the decay of baryonic resonances into $\rho N$ or by $\pi$ annihilation. Observed is a dominance of the $\pi$ annihilations especially in the mass region above the $\phi$ peak. Sub-leading contributions to the $\rho$ tail are found for different $\Delta^*$ and $N^*$ decays, most prominent here is the $\Delta^*(1700)$.

Improved experimental data at the $\phi$ peak will also be valuable to further constrain the $\phi$ production. Previous studies \cite{Staudenmaier:2017vtq, Steinberg:2018jvv} show that the $\phi$ contribution in dielectron invariant mass spectrum is able to aid constraining the branching ratio for various $N^*\rightarrow N\phi$ decays, which is the main production mechanism for $\phi$ mesons in SMASH.

\begin{figure}
	\includegraphics[width=0.95\columnwidth]{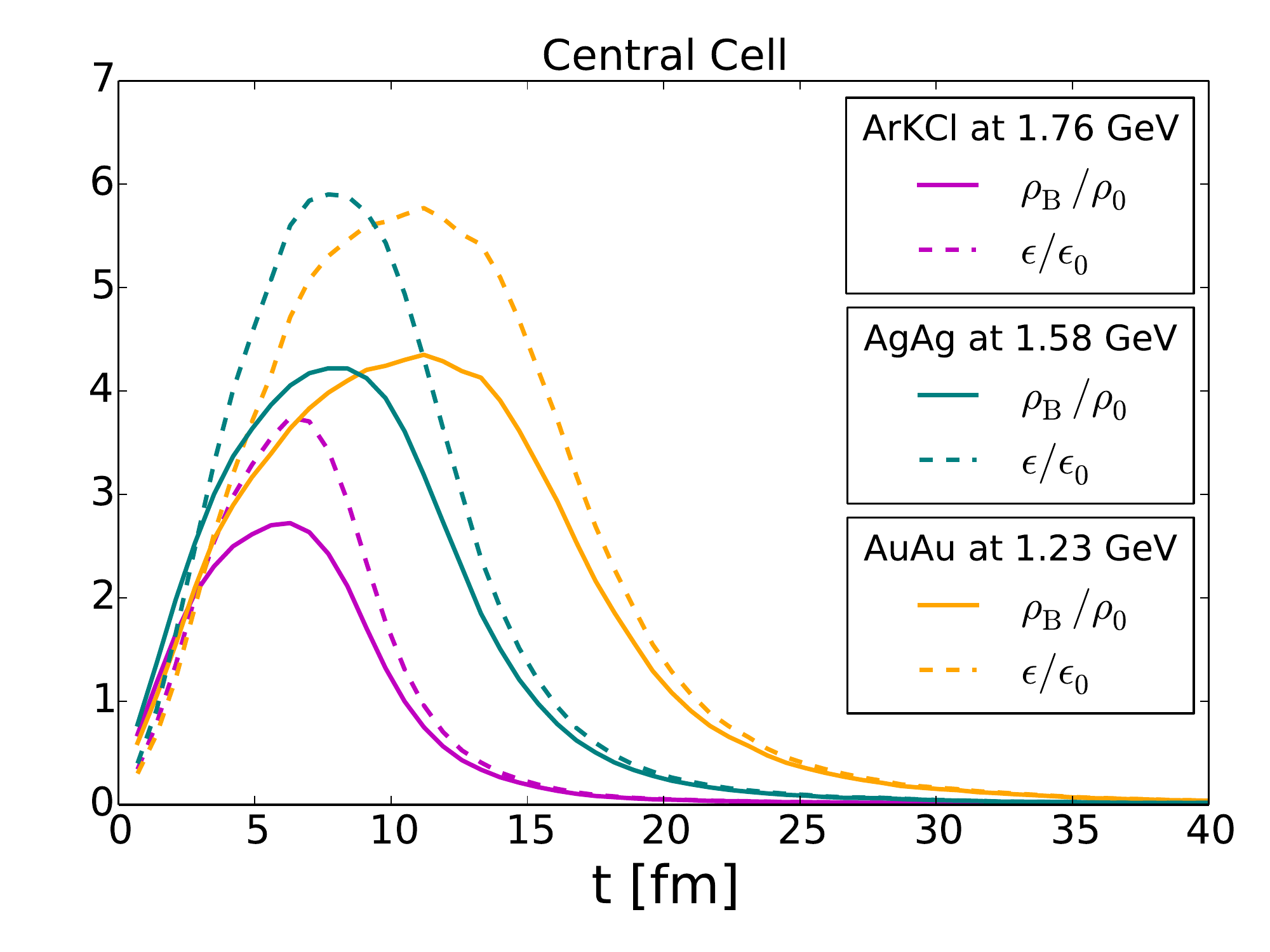}
	\includegraphics[width=0.95\columnwidth]{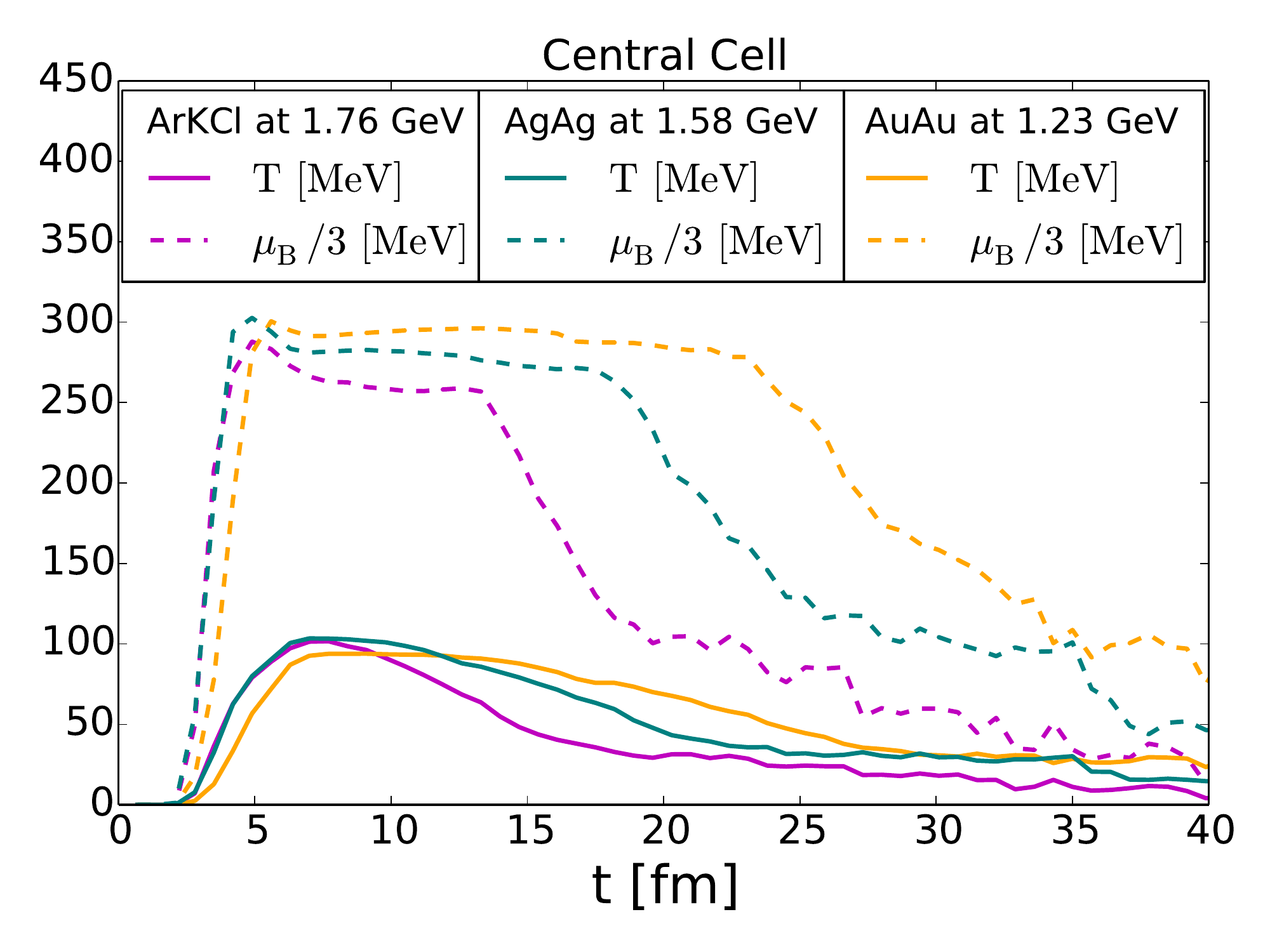}
	\caption{
	    \label{fig:cg-evolution}
		Evolution of the energy and baryon density in units of the ground-state densities $\epsilon_0=146.5\rm{\,MeV/fm^{3}}$ and $\rho_0=0.16\rm{\,fm^{-3}}$ (upper plot),  temperature $T$ and baryochemical potential  $\mu_B$  (lower plot) in the most central cell over time for ArKCl collisions at $E_{\rm{Kin}}=1.76A\,\textrm{GeV}$, AgAg collisions  at $E_{\rm{Kin}}=1.58A\,\textrm{GeV}$ and AuAu collisions at $E_{\rm{Kin}}=1.23A\,\textrm{GeV}$. All results for SMASH-1.6.
	}
\end{figure}

To assess the influence of medium modified spectral functions a coarse-graining approach is applied \cite{Endres:2015fna,Staudenmaier:2017vtq}. The evolution of macroscopic quantities in the central cell in AgAg collisions is shown in Fig.~\ref{fig:cg-evolution} and compared with AuAu and ArKCl reactions (at the beam energies for which experimental data is currently available). The upper plot shows the evolution of the baryon and energy density, the lower one those extracted temperature and baryon chemical potential. The figure nicely illustrates the differences in the evolution of the system at the center of the collision: ArKCl as the smallest system only builds up a smaller density than the larger systems. AgAg builds up the density quicker, since the beam energy is higher, but also falls off faster than AuAu. Both systems, however, again behave similar in terms of maximum density that they reach. AuAu as the largest system maintains large densities the longest. The differences in density mainly translate into differences in the decline of $T$ and $\mu_B$ over time, which shows a clear ordering with the system size.

\begin{figure}
	\includegraphics[width=0.95\columnwidth]{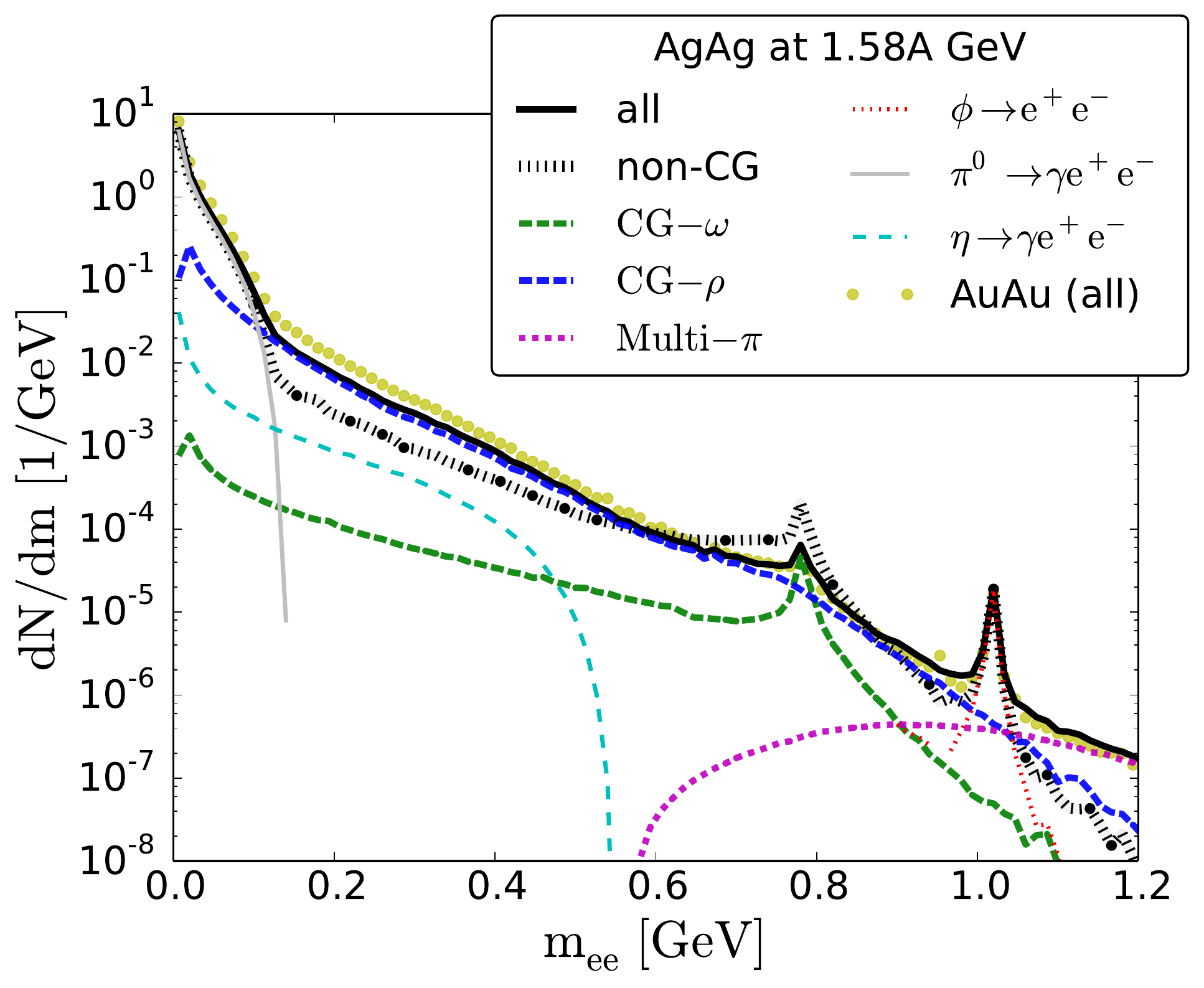}
	\caption{
		 \label{fig:dil-cgAgAg}
		Invariant mass spectrum of dielectrons produced by AgAg collisions at $E_{\rm{Kin}}=1.58A\,\textrm{GeV}$ within the coarse-graining approach. Dashed lines from coarse graining and solid lines from SMASH dilepton production (as in Fig.~\ref{fig:dil-AgAg}).
	}
\end{figure}

The coarse-graining results for the dilepton production in AgAg reactions with $E_{\rm{Kin}}=1.58A\,\textrm{GeV}$ are seen in Fig.~\ref{fig:dil-cgAgAg}. For the contributions of $\pi$, $\eta$ and $\phi$ there is no effect of the medium on the spectral functions and their contributions are taken from the SMASH transport approach (same as in Fig.~\ref{fig:dil-AgAg}). The coarse-graining contributions for the $\rho$ and $\omega$ include the thermal emissions with medium modifications as well as the so called \emph{freeze-out} contributions from space-time regions of the evolution with a low energy density where medium modifications play a negligible role. Also part of the coarse-graining yields is the multi-$\pi$ yield originating from broad multi-$\pi$ states~\cite{vanHees:2006ng}. Fig.~\ref{fig:dil-cgAgAg} reveals that the multi-$\pi$ yield becomes dominant beyond the $\phi$ peak. This essentially is the same finding as above, where the (two-$\pi$) $\rho$ state is the largest contribution (cf. Fig.~\ref{fig:dil-AgAgrho}): the region above the $\phi$ pole mass is dominated by $\pi$ annihilation reactions.

If the total dilepton production from the transport approach alone (\emph{non-CG} in Fig.~\ref{fig:dil-cgAgAg}) is compared to coarse-graining yield, a clear difference is seen. The yield is shifted away from the vector meson peaks. This finding confirms the previously found results for the other large systems ArKCl and AuAu in \cite{Staudenmaier:2017vtq}, which showed that, for larger systems, the dilepton production is sensitive to the inclusion of medium modifications to the vector mesons spectral function already at low beam energies. 

In fact, comparing the total contributions from the coarse-graining approach for AuAu in Fig.~\ref{fig:dil-cgAgAg} to AgAg reveals again (compare Fig.~\ref{fig:dil-AgAg}) that both systems emit dileptons similarly. However, the AuAu spectrum shows a small, but consistent higher emission in the $\rho$ dominated region between $0.15$ and $0.6$ GeV compared to AgAg, which grows towards smaller invariant masses. This hints at even larger medium effects in the larger (AuAu) system.

%###############################################################################
\section{Summary and Outlook} \label{sec:sum}
%###############################################################################

In this work, the particle production of hadrons and dileptons in AgAg collisions at a kinetic energy of $1.58A$ GeV within the SMASH transport approach has been presented allowing for many interesting comparisons to the upcoming experimental data. In addition to the more basic predictions for the multiplicities, rapidity and transverse momentum spectra, special emphasis is put on the study of the production of strange particles. The strange particle multiplicities exhibit the previously discovered $N_{\rm part}$-scaling with system size also for the AgAg system, with a similar exponent than reported by the HADES collaboration. Otherwise, the longitudinal and transverse spectra as well as the multiplicities show the expected behavior and confirm an overall reasonable description of the dynamics. The effective temperatures extracted from transverse mass spectra are on the order of $90-110$ MeV. 

The approach to produce $\Xi$ baryons from heavy $N^*$ resonance decays is able to describe the existing data from HADES with only one free parameter. A branching ratio of $0.5$ for the $N^*\rightarrow\Xi KK$ decays is found to lead to an agreement with the experiment for the $\Xi$ multiplicity and the $\rm \Xi^-/(\Lambda+\Sigma^0)$ ratio in pNb as well as ArKCl. Interestingly, the relation between the $N^*$ decay branching ratios into $\phi$ and $\Xi$ is the same as found in previously studies with other approaches, even though absolutely they are both higher. The comparison of the predictions in this work with future data for AgAg collisions will therefore be able to give further insights into the origin of the $\phi$ and $\Xi$ production mechanism. Even though numerically challenging for the rarely produced $\phi$ and $\Xi$, these results could also be utilized to study the role of potentials on the production mechanism in the future.

The dielectron production is found to be overall very similar to the previously studied AuAu system suggesting that the dilepton production behaves the same for smaller systems with higher energies compared to larger systems but with lower beam energy. The main source for the dilepton emission for invariant masses higher than the $\phi$ pole mass, is the $\rho$ contribution, which originates from (two) $\pi$ annihilations. Decays of baryonic resonances are only sub-leading. Employing a coarse-graining approach, this finding is confirmed by also identifying the multi-$\pi$ contribution as dominant beyond the $\phi$ peak. The employed coarse-graining approach furthermore allows to gauge that the invariant mass spectrum is sensitive to medium modifications of the spectral function of $\rho$ and $\omega$. This confirms previous results for ArKCl and AuAu, which also showed that medium modifications for larger systems are already relevant at the discussed low beam energies.

%###############################################################################
\begin{acknowledgements}
We acknowledge R. Rapp for providing the parameterization of the spectral functions for the coarse-graining approach. J.S. is funded by the Deutsche Forschungsgemeinschaft (DFG, German Research Foundation) – Project number 315477589 – TRR 211. Computational resources have been provided by the Center for Scientific Computing (CSC) at the Goethe-University of Frankfurt and the GreenCube at GSI.
\end{acknowledgements}
%###############################################################################

%-------------------------------------------------------------------------------
\bibliography{AgAg_paper}
%-------------------------------------------------------------------------------

%###############################################################################
\appendix

%-------------------------------------------------------------------------------
\section{Definition of Spectators}
%-------------------------------------------------------------------------------
\label{appendix_spec}

\begin{figure}
	\includegraphics[width=0.95\columnwidth]{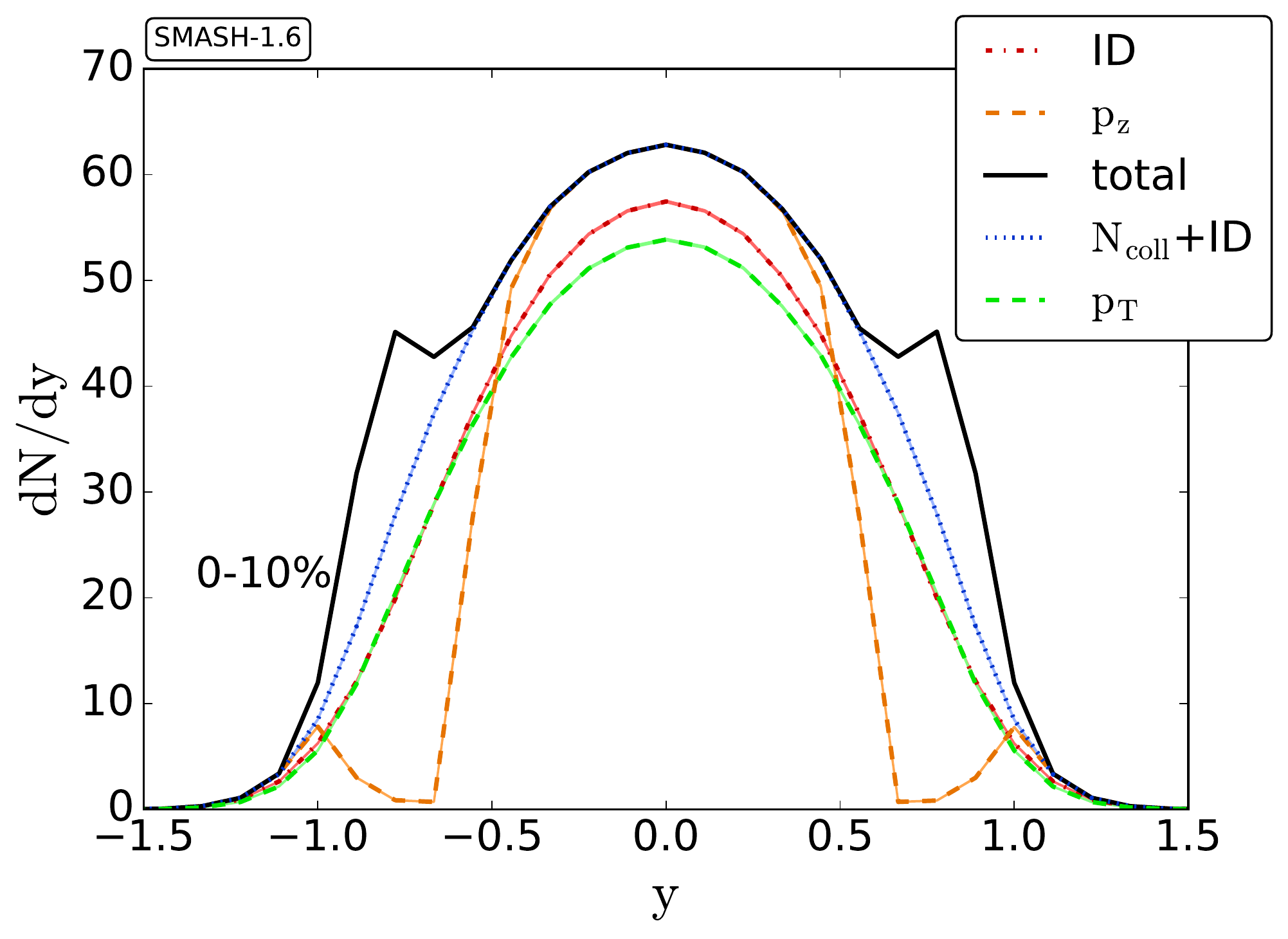}\caption{\label{fig:specs1}Proton rapidity distributions for different applied cutting criteria for the elimination of spectators in AgAg collisions at $E_{\rm Kin}=1.58A$ GeV.}
\end{figure}

The definition of participating nucleons and spectators is crucial for the understanding and analysis of heavy-ion collisions since it is directly connected to the concept of centrality and the collision geometry. Here, four different ways of defining participants in SMASH are introduced. The results of the different cutting criteria are shown in the rapidity distributions in Fig. \ref{fig:specs1} in AgAg collisions at $E_{\rm Kin}=1.58A$ GeV.

One way to eliminate spectators is by using the ID number (red, dashed line) that is assigned to every particle in ascending order when they are initialized (starting with 0). In this case, only nucleons that interact inelastically are defined as participants, and all other nucleons with ID$<$2A present in the final state are considered spectators. For the $p_z$ criterion (orange, dashed line) nucleons with $p_z$ within the range of initial momenta in z-direction are defined as spectators and are cut. In a similar fashions, the $p_T$ cut (green, dashed line) eliminates nucleons with $p_T$ within the range of initial transverse momentum. Lastly, the $N_{\rm coll}$+ID cut is introduced (blue, dotted line). It modifies the ID cut by also classifying elastically scattered nucleons as participants which is characterized by $N_{\rm coll}>0$. In this work, the $N_{\rm coll}$+ID cut is used for all spectra.

%-------------------------------------------------------------------------------
\section{Definition of Centrality Classes}
%-------------------------------------------------------------------------------
\label{appendix_centrality}

As shown in Table \ref{tab:c_cent}, the centrality classes in this work are mapped onto certain impact parameter ranges and numbers of participants using Glauber calculations provided by \cite{Miskowiec} for AuAu, AgAg, ArKCl and CC collisions at $E_{\rm Kin}=1.58A$ GeV.

\begin{table*}
    \begin{tabularx}{5.3cm}{XXX}
		\hline
		\hline
		\textbf{AuAu} & & \\
		\hline
		$C$ & $b$ [fm]& $\rm N_{part}$\\
		\hline
		0-10\%  & 0.0-4.7   & 307.4\\
		10-20\% & 4.7-6.6 & 210\\
		20-30\% & 6.6-8.1 & 148.6\\
		30-40\%	&8.1-9.3  & 97.9\\
		\hline
		\hline
	\end{tabularx}%
	\vspace{0.2cm} 
	
    \begin{tabularx}{5.3cm}{XXX}
	    \hline
	    \hline
		\textbf{AgAg} &  & \\
		\hline
		$C$ & $b$ [fm]& $\rm N_{part}$\\
		\hline
		0-10\%  & 0.0-3.8   & 168.1\\
		10-20\% & 3.8-5.4 & 112\\
		20-30\% & 5.4-6.7 & 75.8\\
		30-40\%	& 6.7-7.7  & 53.4\\
		\hline
		\hline
	\end{tabularx}%
	\vspace{0.2cm}

    \begin{tabularx}{5.3cm}{XXX}
    	\hline
    	\hline
    	\textbf{ArKCl} &  & \\
    	\hline
    	$C$ & $b$ [fm]& $\rm N_{part}$\\
    	\hline
    	0-10\%  & 0.0-2.7   & 53.8\\
    	10-20\% & 2.7-3.8 & 39.9\\
    	20-30\% & 3.8-4.7 & 25\\
    	30-40\%	& 4.7-5.4  & 18.4\\
    	\hline
    	\hline
    \end{tabularx}%
    \vspace{0.2cm} 
    
    \begin{tabularx}{5.3cm}{XXX}
    	\hline
    	\hline
    	\textbf{CC} & & \\
    	\hline
    	$C$ & $b$ [fm]& $\rm N_{part}$\\
    	\hline
    	0-10\%  & 0.0-1.7   & 13.1\\
    	10-20\% & 1.7-2.4 & 11.5\\
    	20-30\% & 2.4-3.0 & 7.8\\
    	30-40\%	& 3.0-3.4  & 5.7\\
    	\hline
    	\hline
    \end{tabularx}
	\caption{\label{tab:c_cent}Centrality Classes $C$ with corresponding impact parameter $b$ and number of participants $\rm N_{part}$ at $E_{\rm Kin}=1.58A$ GeV. These values result from Glauber calculations performed with \cite{Miskowiec}.}
\end{table*}

%-------------------------------------------------------------------------------
\section{Transverse Mass Spectra}
\begin{figure*}
	\includegraphics[width=0.45\textwidth]{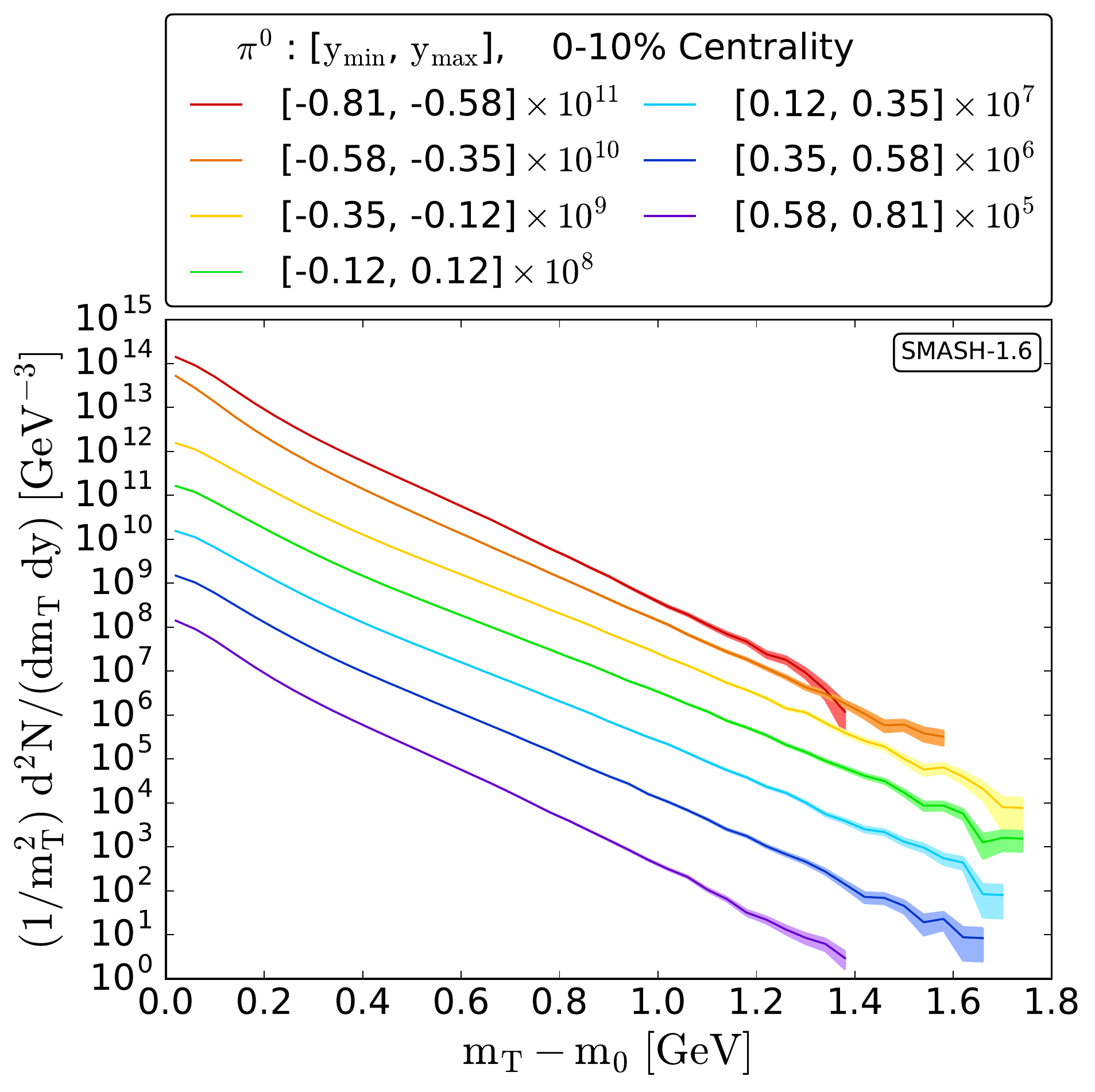}
	\includegraphics[width=0.45\textwidth]{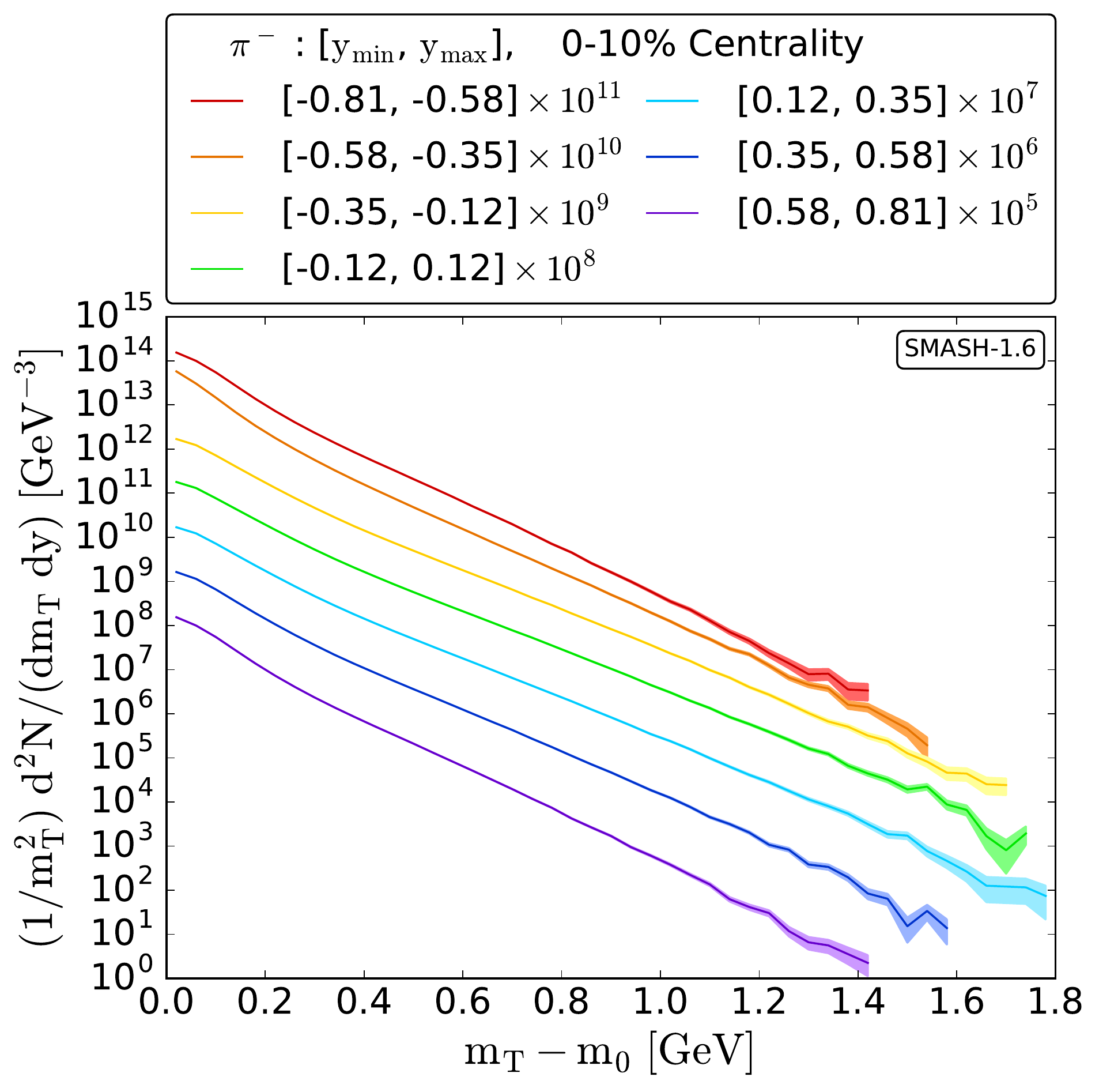} 
	\\
	\includegraphics[width=0.45\textwidth]{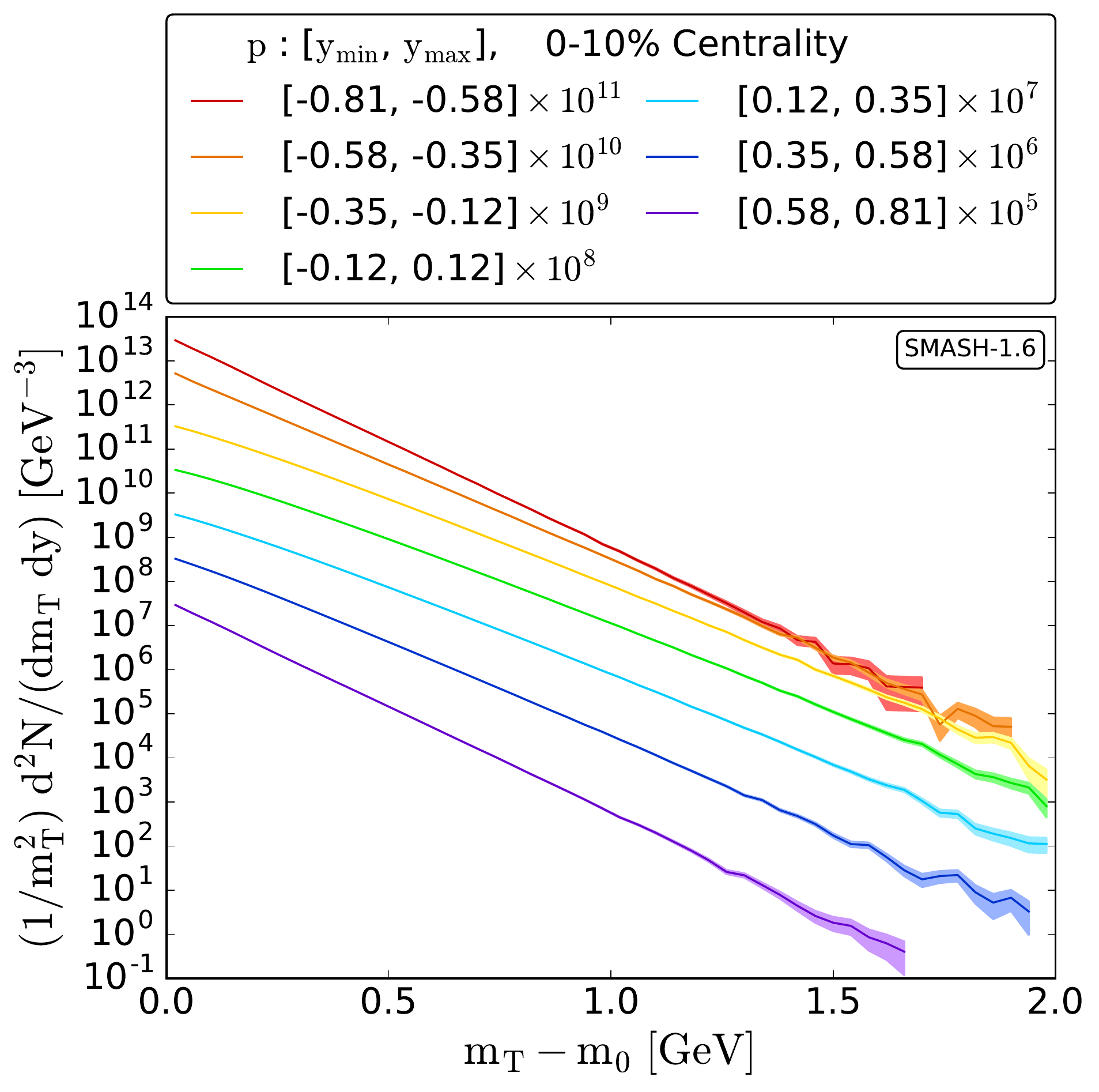}
	\includegraphics[width=0.45\textwidth]{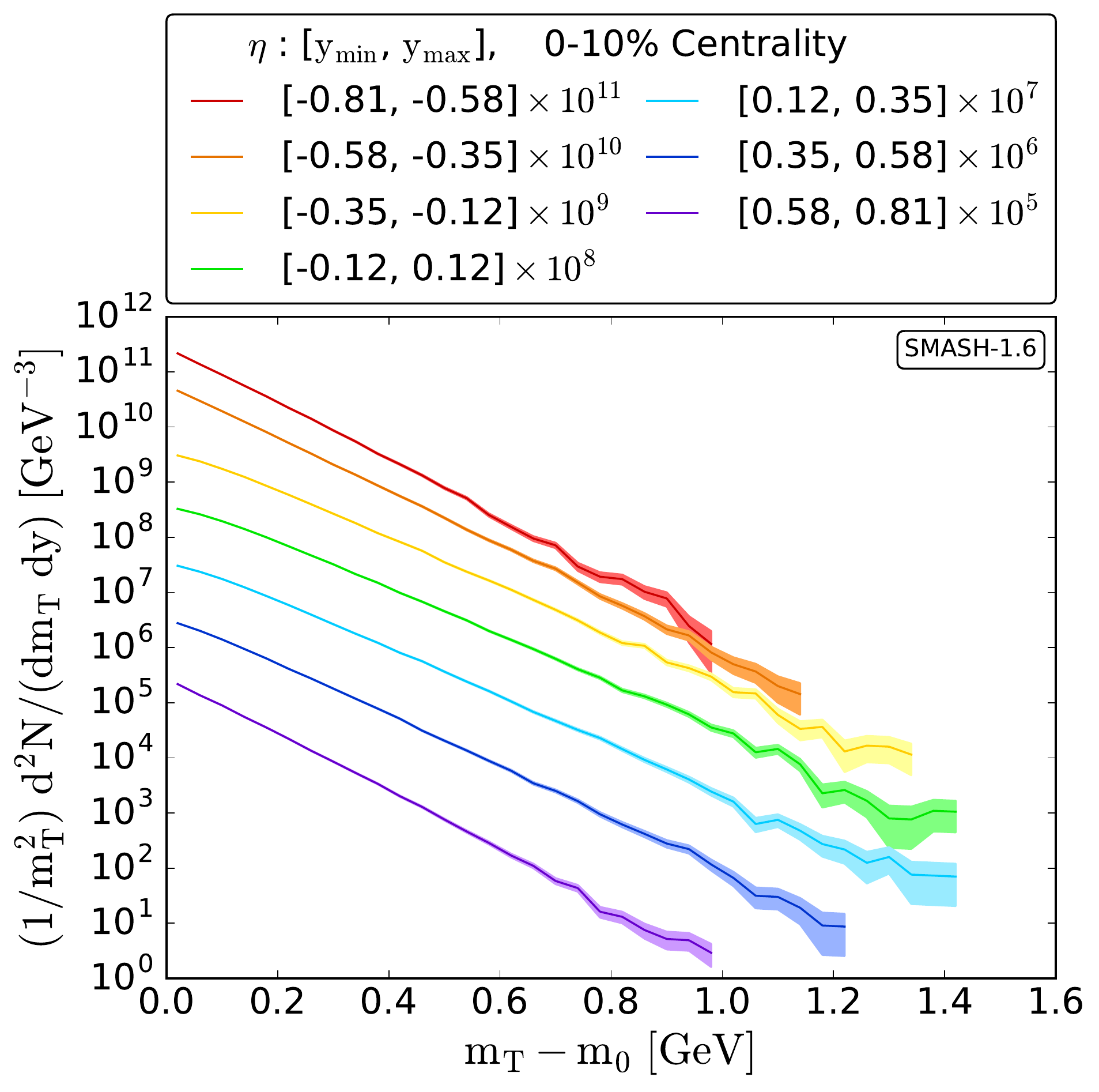}
	\caption{\label{fig:app-mt}Mean transverse mass distribution of the $\pi^0$, $\pi^-$, $\eta$ and $p$  in AgAg collisions at $E_{\rm Kin}=1.58A$ GeV.}

\end{figure*}

\begin{figure*}

\includegraphics[width=0.95\columnwidth]{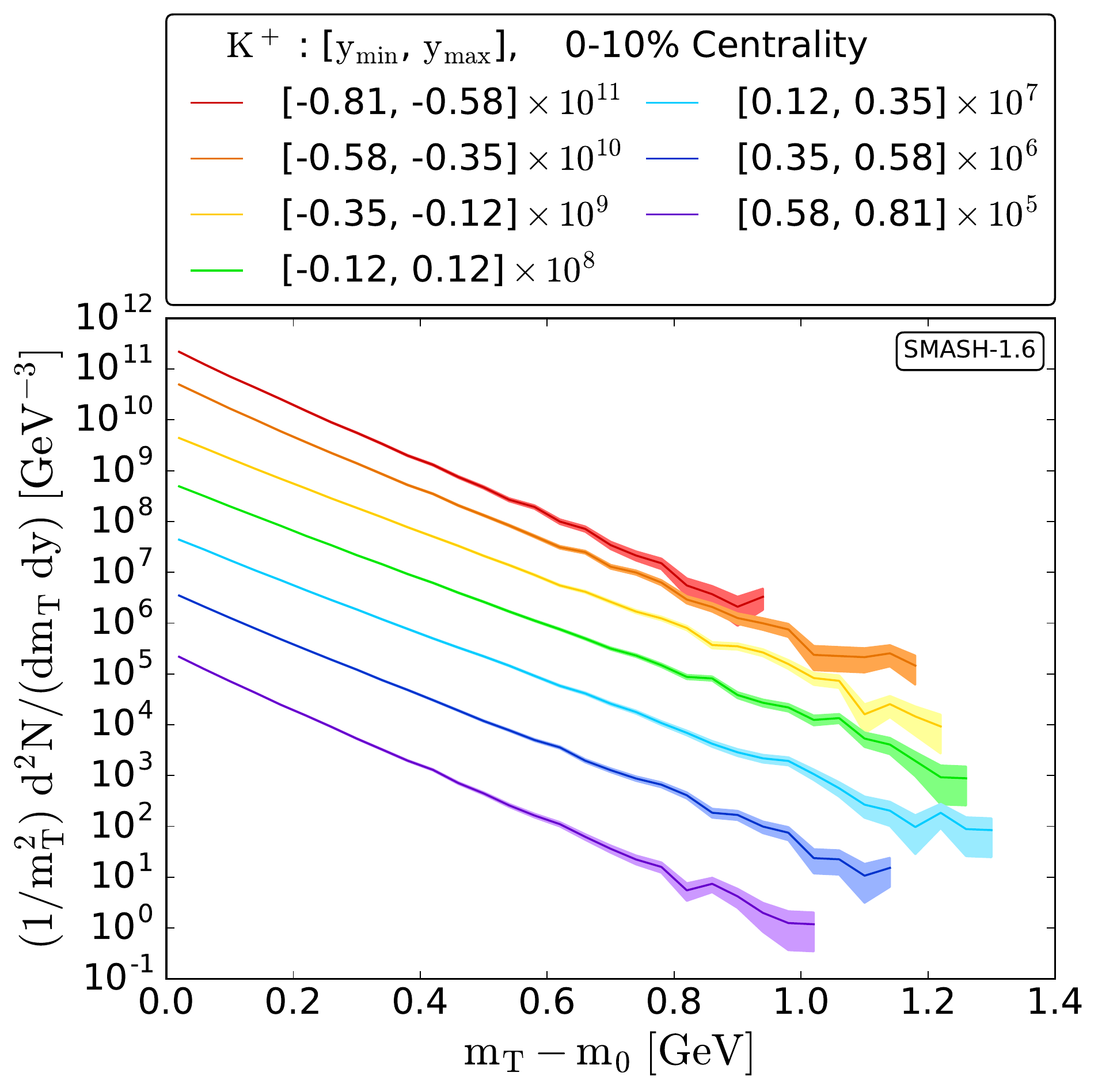}
\includegraphics[width=0.95\columnwidth]{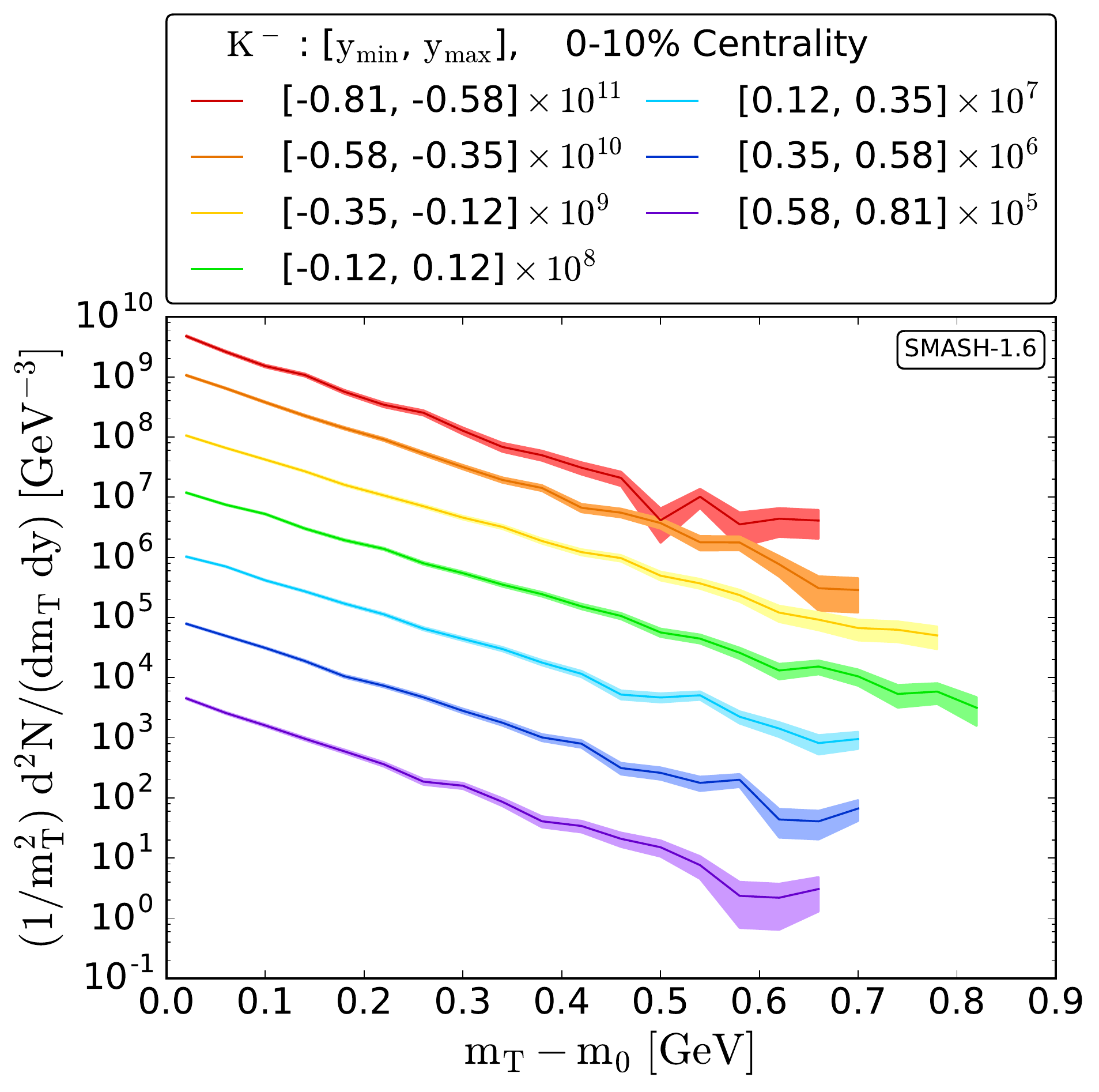}
\includegraphics[width=0.95\columnwidth]{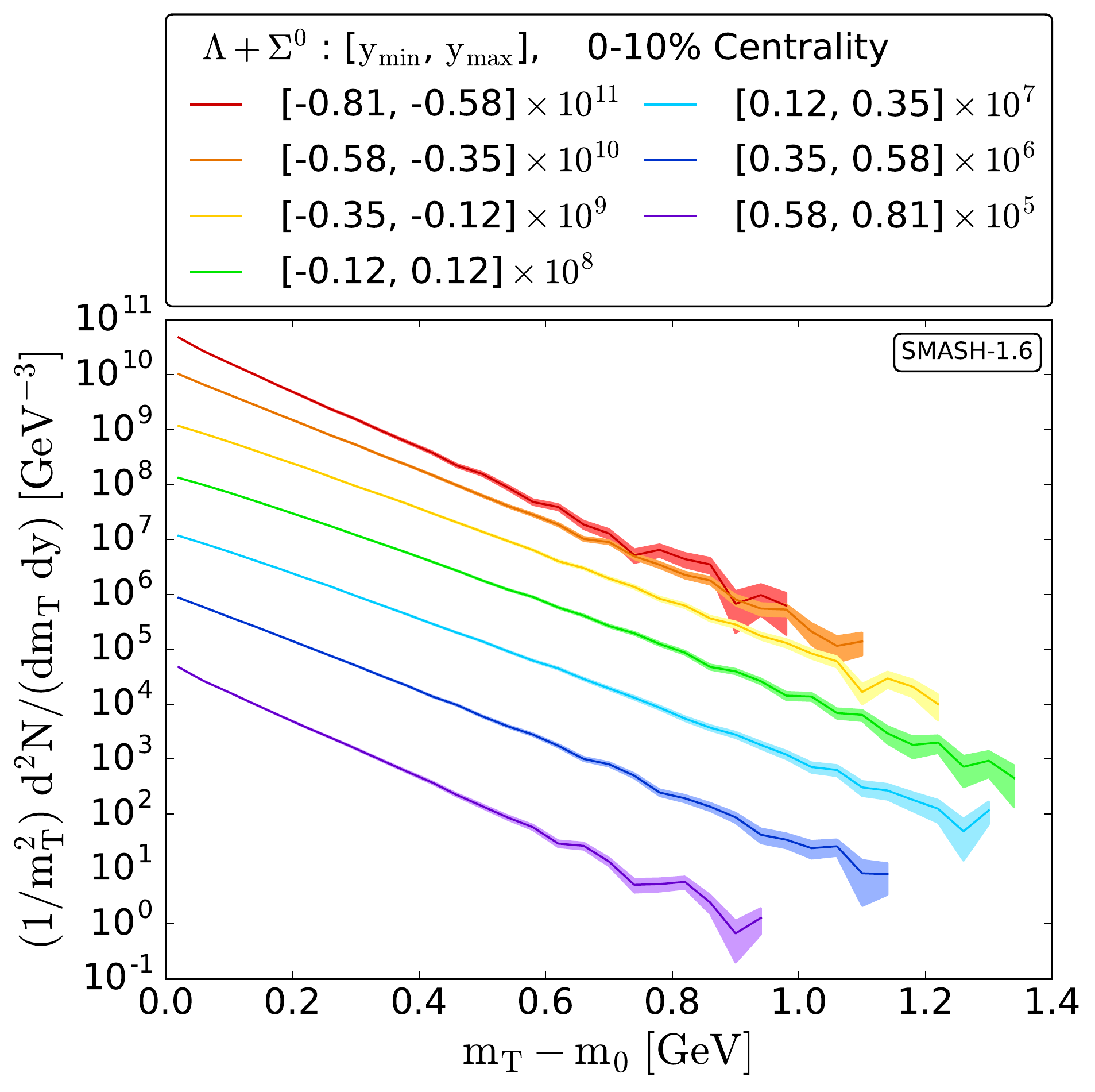}

\caption{\label{fig:app-mt-2}Mean transverse mass distribution of K$^-$, K$^+$ and $\Lambda$ in AgAg collisions at $E_{\rm Kin}=1.58A$ GeV. }

\end{figure*}

%-------------------------------------------------------------------------------
\label{appendix_transverse_mass}
The transverse mass distributions of the  $\pi^0$, $\pi^-$, $\eta$ and $p$ are depicted in Fig. \ref{fig:app-mt} and the ones of the K$^-$, K$^+$ and $\Lambda$  are depicted in Fig. \ref{fig:app-mt-2} for several rapidity bins.   

\end{document}